\documentclass{aa}
\usepackage{txfonts}
\usepackage{graphicx}
\usepackage{natbib}
\bibpunct{(}{)}{;}{a}{}{,}

\begin{document}

\newcommand{\be}{\begin{equation}}
\newcommand{\ee}{\end{equation}}
\newcommand{\bdm}{\begin{displaymath}}
\newcommand{\edm}{\end{displaymath}}
\newcommand{\bea}{\begin{eqnarray}}
\newcommand{\eea}{\end{eqnarray}}

\title{ECHO: an Eulerian Conservative High Order scheme for general relativistic magnetohydrodynamics and magnetodynamics}

\author{
        L. Del Zanna    \inst{1}
\and    O. Zanotti      \inst{1}
\and    N. Bucciantini  \inst{2} 
\and    P. Londrillo    \inst{3}
}

\institute{
Dipartimento di Astronomia e Scienza dello Spazio,
Universit\`a di Firenze, Largo E. Fermi 2, 50125 Firenze, Italy
\\ \email{ldz@arcetri.astro.it}
\and
Astronomy Department, University of California at Berkeley,
601 Campbell Hall, Berkeley, CA 94720-3411, USA
\and
INAF - Osservatorio Astronomico di Bologna, 
Via C. Ranzani 1, 40127 Bologna, Italy
}

\date{Received ...; accepted ...}

\authorrunning{L.~Del Zanna et al.}
\titlerunning{ECHO: an Eulerian Conservative High Order scheme for GRMHD and GRMD}

\abstract
{}
{We present a new numerical code, ECHO, based on an \emph{Eulerian Conservative High Order} scheme for time dependent three-dimensional general relativistic magnetohydrodynamics (GRMHD) and magnetodynamics (GRMD). ECHO is aimed at providing a shock-capturing conservative method able to work at an arbitrary level of formal accuracy (for smooth flows), where the other existing GRMHD and GRMD schemes yield an overall second order at most. Moreover, our goal is to present a general framework, based on the $3+1$ Eulerian formalism, allowing for different sets of equations, different algorithms, and working in a generic space-time metric, so that ECHO may be easily coupled to any solver for Einstein's equations.}
{Our finite difference conservative scheme previously developed for special relativistic hydrodynamics and MHD is here extended to the general relativistic case. Various high order reconstruction methods are implemented and a two-wave approximate Riemann solver is used. The induction equation is treated by adopting the \emph{Upwind Constrained Transport} (UCT) procedures, appropriate to preserve the divergence-free condition of the magnetic field in shock-capturing methods. The limiting case of magnetodynamics (also known as force-free degenerate electrodynamics) is implemented by simply replacing the fluid velocity with the electromagnetic drift velocity and by neglecting the matter contribution to the stress tensor.}
{ECHO is particularly accurate, efficient, versatile, and robust. It has been tested against several astrophysical applications, like magnetized accretion onto black holes and constant angular momentum thick disks threaded by toroidal fields. A novel test on the propagation of \emph{large amplitude} circularly polarized Alfv\'en waves is proposed and this allows us to prove the spatial and temporal high order properties of ECHO very accurately. In particular, we show that reconstruction based on a \emph{Monotonicity Preserving} filter applied to a fixed 5-point stencil gives highly accurate results for smooth solutions, both in flat and curved metric (up to the nominal fifth order), while at the same time providing sharp profiles in tests involving discontinuities.}
{}


\maketitle


\section{Introduction}
\label{sect:intro}

Compact objects like black holes and neutron stars interacting with the relativistic plasma in the surrounding regions are believed to be responsible for many of high energy phenomena in astrophysics. The most luminous sources, namely active galactic nuclei or gamma-ray bursts, are likely to be powered by the conversion of gravitational energy of rotating black holes into electromagnetic fields and a plasma of relativistic particles \citep{blandford77}. A similar mechanism had been previously proposed to generate the magnetospheric plasma and ultimately a Poynting flux dominated wind from rotating neutron stars \citep{goldreich69}. The presence of the magnetic field is crucial in all the situations outlined above. The magnetic field could also be important in the phases of gravitational collapse that then give rise to the compact objects themselves, because the freeze-in condition valid for highly conducting plasmas would allow an initially negligible field to be enhanced by the collapse to such high intensities to be ultimately dominant. The physical frameworks in which these mechanisms are treated are usually that of general relativistic magnetohydrodynamics (GRMHD) or, when the electromagnetic field contribution is dominant over the matter contribution, that of force-free degenerate electrodynamics \citep{komissarov02,komissarov04a}, also known as \emph{magnetodynamics} \citep[GRMD,][]{komissarov06c}. In both cases the electromagnetic fields interact strongly with the plasma, in such a way that freely moving charges are supposed to screen efficiently any local electric field and to maintain quasi-neutrality.

A great impulse to the study of these complex phenomena has come from numerical simulations, especially in the last decade. Since relativistic magnetized flows are often associated with the formation of strong shocks and different kinds of discontinuities, it is thanks to the development of conservative shock-capturing, or Godunov-type, methods that this progress has been possible. After the first applications to special and general relativistic hydrodynamics \citep[e.g.][]{font94,eulderink94,banyuls97,aloy99}, see also \citet{marti03,font03} for reviews, \citet{komissarov99} first proposed a multi-dimensional shock-capturing code for special relativistic MHD (RMHD). These schemes are all based on the so-called Roe-type methods, widely used in computational gas dynamics, in which the solution of the local Riemann problem at any cell interface is constructed by means of a full decomposition into characteristic waves. However, while this approach is perfectly feasible for purely hydrodynamic flows, in RMHD the spectral structure of the system is much harder to resolve, due to the increase in number (from five to seven) and complexity (eigenvalues are to be found numerically) of the characteristic waves, and to the presence of a preferential direction that may lead to non-strict hyperbolicity of the local system. Furthermore, the solenoidal constraint for the magnetic field in multi-dimensions requires a special numerical treatment, which must be compatible with the conservative approach.

Within the family of shock-capturing conservative schemes, a different strategy was followed in our previous investigations on numerical relativistic hydrodynamics \citep{delzanna02}, hereafter Paper~I, and MHD \citep{delzanna03}, hereafter Paper~II, relying on the promising results obtained for classical MHD \citep{londrillo00}. As shown in these works, accurate and robust schemes can be devised even if the characteristic spectral decomposition of the equations is not fully known, or exploited, because this lack of knowledge is somehow compensated by resorting to higher (third) order reconstruction of intercell variables, leading to a more accurate setup of the local Riemann problem. By doing so, even simple one or two-wave approximate Riemann solvers (also known as central-type schemes) are capable of resolving all kinds of structures, thus avoiding the problems related to the complexity in spectral decomposition at the price of a slightly higher numerical diffusion of contact and Alfv\'enic discontinuities. Many other shock-capturing numerical codes for RHMD and GRMHD (some of them even with evolving space-time metric) share the same philosophy of a simplified Riemann solver \citep{gammie03,duez05,shibata05,leismann05,mignone06,anton06}, though all of them are based on finite difference or finite volume second order schemes. In \citet{anton06} an RMHD Roe solver is also used in some tests, via a local coordinate transformation to flat metric \citep{pons98}. Moreover, different methods other than Godunov-type have also been proposed for GRMHD \citep{koide99,koide03,devilliers03a,anninos05} and (GR)MD \citep{spitkovsky06}. See also the reviews by \citet{wilson03,font03}.

These codes have been extensively applied to many astrophysical situations involving relativistic plasmas and compact objects. Relevant examples of these applications include the validation of the Blandford-Znajek mechanism for the extraction of rotational energy from a Kerr black hole \citep{komissarov01,koide03,komissarov04a,mckinney04,komissarov05,mckinney05}; the spin evolution of a black hole  under the effect of different physical processes \citep{gammie04}; the problem of jet formation in a black hole-accretion disk system \citep{koide00,devilliers03b,mizuno04,nishikawa05,devilliers05,mckinney06b,hawley06,koide06}; the time evolution of a neutron star magnetosphere, both in the MHD regime \citep{komissarov06a} and in the force-free approximation \citep{mckinney06c,spitkovsky06}; the acceleration of magnetized pulsar winds \citep{bucciantini06} and the dynamics and emission properties of their related nebulae \citep{komissarov04b,delzanna04,bucciantini05,delzanna06}; the morphology and the dynamics of axisymmetric relativistic jets with different magnetic field topologies \citep{leismann05}; the collapse, in full general relativity of a hyper-massive neutron star \citep{shibata06,duez06a}, also including the effects of differential rotation \citep{duez06b}. All of these applications, that do not pretend to provide a complete list, surely give a  sample of the fundamental contributions that numerical simulations have been offering to our understanding of the highly complex physical processes induced by the relativistic plasma around compact objects.

In this paper we present the main features of our new GRMHD code ECHO, based on an \emph{Eulerian Conservative High Order} scheme, that completes and refines our previous works for special relativity (Paper~I and II). The issue of high numerical accuracy in conservative schemes becomes of great importance when not only shocks and discontinuities, but also fine smooth structures like turbulent fields and waves, are of primary interest. These small scale structures can be smeared out by the excessive numerical diffusion typical of low order schemes. Furthermore, higher than second order accuracy is desirable when moving to 3-D, where numerical grids are necessarily limited in size. This specially applies to GR, due to the gradients of the metric terms that must be treated with appropriate resolution. High order schemes are commonly used in classical gas dynamics \citep[e.g.][]{shu97}, and the general recipes to apply these methods to MHD were given in \citet{londrillo00,londrillo04}, where the solenoidal constraint for the magnetic field was enforced as a built-in condition (\emph{Upwind Constrained Transport} method, UCT). Here we extend this framework to GRMHD by taking advantage of the formalism for the $3+1$ splitting of space-time \citep[e.g.][]{thorne82}. Specifically, we write all terms entering the conservative form of the GRMHD equations as quantities measured by the so-called \emph{Eulerian} observer associated with the three-dimensional metric (not necessarily diagonal), highlighting the closest possible comparison with the equations of MHD and RMHD by using three-dimensional vectors and tensors alone. As a consequence, we are able to write the source terms in such a way that they do not contain four-dimensional Christoffel symbols explicitly, and are therefore very easy to implement numerically. We then incorporate in the $3+1$ formalism the modifications proposed by \citet{mckinney06a} to allow a GRMHD code to solve the equations in the force-free limit of magnetodynamics (GRMD).

The plan of the paper is as follows. In Sect.~\ref{sect:eqs} we present the $3+1$ form of the GRMHD equations. Sect.~\ref{sect:echo} contains a description of the essential features of our numerical scheme. Sects.~\ref{sect:grmhd_tests} and \ref{sect:grmd_tests} are devoted to a presentation of the most important numerical tests performed in GRMHD and GRMD, respectively. Finally, the conclusions are reported in Sect.~\ref{sect:concl}. In the following we will assume a signature $\{-,+,+,+\}$ for the space-time metric and we will use Greek letters $\mu,\nu,\lambda,\ldots$ (running from 0 to 3) for four-dimensional space-time tensor components, while Latin letters $i,j,k,\ldots$ (running from 1 to 3) will be employed for three-dimensional spatial tensor components. Moreover, we set $c=G=M_{\sun}=1$ and make use of the Lorentz-Heaviside notation for the electromagnetic quantities, thus all $\sqrt{4\pi}$ factors disappear.

\section{GRMHD equations in $3+1$ conservative form}
\label{sect:eqs}

\subsection{Covariant approach}
\label{sect:cov}

We start with a brief presentation of the GRMHD equations in covariant form. Standard derivations of the laws of fluid dynamics and electrodynamics in covariant form may be found in books such as \citet{landau62,weinberg72,misner73}, while for the MHD equations and their basic properties see \citet{lichnerowicz67,anile89}. Consider an ideal fluid interacting with an electromagnetic field. The corresponding Euler equations are
\be
\nabla_{\mu} (\rho u^{\,\mu})=0,
\label{eq:mass}
\ee
\be
\nabla_{\mu}T^{\mu\nu}=0,
\label{eq:momentum}
\ee
where $\nabla_{\mu}$ is the space-time covariant derivative. Eq.~(\ref{eq:mass}) is the usual mass conservation law, in which $\rho$ is the mass density as measured in the (Lagrangian) frame comoving with the fluid four-velocity $u^{\,\mu}$. Eq.~(\ref{eq:momentum}) is the law of momentum-energy conservation, where the total momentum-energy tensor is made up by two contributions, $T^{\mu\nu}=T^{\mu\nu}_{m}+T^{\mu\nu}_{f}$, one due to matter
\be
T^{\mu\nu}_{m}=\rho h\,u^{\,\mu}u^{\nu}+pg^{\,\mu\nu},
\label{eq:T_matter}
\ee
and the other due to the electromagnetic field
\be
T^{\mu\nu}_{f}={F^{\mu}}_{\lambda}F^{\nu\lambda}-\textstyle{\frac{1}{4}}(F^{\lambda\kappa}F_{\lambda\kappa})g^{\,\mu\nu}.
\label{eq:T_field}
\ee
In the above expressions $g^{\,\mu\nu}$ is the space-time metric tensor, $h=1+\epsilon + p/\rho$ is the specific enthalpy (including rest mass energy contribution), $\epsilon$ is the specific internal energy, $p$ is the thermal pressure, $F^{\mu\nu}$ is the electromagnetic field (antisymmetric) tensor. When considered separately, the two components of the stress tensor are not conserved
\be
\nabla_{\mu}T^{\mu\nu}_m=-\nabla_{\mu}T^{\mu\nu}_f=-J_{\mu}F^{\mu\nu},
\label{eq:lorentz}
\ee
where $J^{\,\mu}$ is the four-vector of current density and the last term is the electromagnetic force acting on the conducting fluid. The fields obey the two Maxwell equations
\be
\nabla_{\mu}F^{\mu\nu} = -J^{\nu},
\label{eq:maxwell1}
\ee
\be
\nabla_{\mu}F^{*\mu\nu} = 0, 
\label{eq:maxwell2}
\ee
where $F^{*\mu\nu}=\frac{1}{2}\epsilon^{\,\mu\nu\lambda\kappa}F_{\lambda\kappa}$ is the dual of the electromagnetic tensor, and $\epsilon^{\,\mu\nu\lambda\kappa}$ is the space-time Levi-Civita tensor density, that is $\epsilon^{\,\mu\nu\lambda\kappa}=(-g)^{-1/2}[\mu\nu\lambda\kappa]$ (and $\epsilon_{\mu\nu\lambda\kappa}=-(-g)^{1/2}[\mu\nu\lambda\kappa]$), with $g=\mathrm{det}\{g_{\mu\nu}\}$ and $[\mu\nu\lambda\kappa]$ is the alternating Levi-Civita symbol.

Since we are dealing with a (perfectly) conducting fluid, a general relativistic extension of (ideal) Ohm's law is needed. This translates in a condition of vanishing electric field in the comoving frame
\be
F^{\mu\nu}u_{\nu}=0.
\label{eq:mhd}
\ee
From a physical point of view it means that the freely moving charges in a plasma are supposed to be always able to screen any electric field that may arise locally. The extra condition imposed on $F^{\mu\nu}$ in Eq.~(\ref{eq:mhd}) makes the first Maxwell equation redundant, and Eq.~(\ref{eq:maxwell1}) is only needed to calculate the four-current $J^{\,\mu}$, which is now a derived quantity like in non-relativistic MHD. The system of GRMHD equations is then closed by choosing an equation of state (EoS) $p=p(\rho,\epsilon)$. Different relativistic EoS may be employed, and thus we will leave it unspecified in our formulation. However, all numerical tests presented here will make use of the standard $\gamma$-law for a perfect gas
\be
p(\rho,\epsilon)=(\gamma-1)\,\rho\,\epsilon \Rightarrow h=1 + \frac{\gamma}{\gamma-1}\frac{p}{\rho},
\label{eq:eos}
\ee
with $\gamma=5/3$ for a non-relativistic fluid and $\gamma=4/3$ when $p\gg\rho$ ($\rho h \to 4p$). Finally, note that for an \emph{ideal} fluid (thus in the absence of shocks or other sources of dissipation) the total energy conservation law is equivalent to the adiabatic equation
\be
u^{\,\mu}\nabla_{\mu} s=0\Rightarrow \nabla_{\mu}(\rho s u^{\,\mu})=0,
\label{eq:adiabatic}
\ee
even in the GRMHD case \citep[e.g.][]{anile89}. Here $s$ is any function of the specific entropy (in the comoving frame), and in the case of a fluid with a $\gamma$-law EoS we can take $s=p/\rho^\gamma$.

\subsection{The $3+1$ splitting of space-time}
\label{sect:3+1}

In spite of their elegant and compact form, the GRMHD covariant equations described above are not suitable for numerical integration, where the temporal coordinate must be clearly singled out. The most widely used formalism is that based on the so-called $3+1$ decomposition of the equations. For a comprehensive treatment and references the reader is referred to \citet{thorne82}, or, for a more recent work, see \citet{baumgarte03}. 

In the $3+1$ formalism, the four-dimensional space-time is foliated into non-intersecting space-like hyper-surfaces $\Sigma_t$, defined as iso-surfaces of a scalar time function $t$. Let then
\be
n_{\mu}=-\alpha\nabla_{\mu} t,~~~~  (n_{\,\mu} n^{\,\mu}=-1)
\label{eq:n}
\ee
be the future-pointing time-like unit vector normal to the slices $\Sigma_t$, where $\alpha$ is called the \emph{lapse function}. The observer moving with four-velocity $n^{\,\mu}$ is called \emph{Eulerian} \citep{smarr78}, and all quantities may be decomposed in the corresponding frame. Thus, any vector $V^{\,\mu}$ (or similarly a tensor) may be projected in its temporal component $V^{\hat{n}}=-n_{\mu}V^{\,\mu}$ and spatial component $\perp V^{\,\mu}=(g^{\,\mu}_{\nu}+n^{\,\mu}n_{\nu})V^{\nu}$. In particular, a three-dimensional spatial metric $\gamma_{\mu\nu}$ can be induced on $\Sigma_t$ by the four-dimensional metric. Application of the projection operator gives
\be
\gamma_{\mu\nu}=\perp g_{\mu\nu}=g_{\mu\nu}+n_{\mu} n_{\nu},
\label{eq:gamma}
\ee
so that we can also identify $\perp\equiv\perp^{\mu}_{\nu}=\gamma^{\mu}_{\nu}$. At this point, it is convenient to introduce a coordinate system $x^{\,\mu}=(t,x^i)$ adapted to the foliation $\Sigma_t$. The line element is usually given in the so-called ADM \citep{arnowitt62} form:
\be
\mathrm{d}s^2 = \! -\alpha^2\mathrm{d}t^2+\gamma_{ij}\,(\mathrm{d}x^i\!+\beta^i\mathrm{d}t)(\mathrm{d}x^j\!+\beta^j\mathrm{d}t),
\label{eq:adm}
\ee
where $\beta^{\,\mu}$ is called \emph{shift vector}, an arbitrary spatial vector ($\beta^{\,\mu}n_{\mu}=0$). Notice that the spatial metric $\gamma_{ij}$ can now be used for the raising and lowering of indices for purely spatial vectors and tensors. In this coordinate system the unit vector components are
\be
n_{\mu}=(-\alpha,0_i),~~~~ n^{\,\mu}=(1/\alpha,-\,\beta^i/\alpha),
\ee
and any spatial vector $V^\mu$ (or tensor) must necessarily have a vanishing contravariant temporal component $V^t=0$, whereas its covariant temporal component is $V_t=g_{\mu t}V^\mu=\beta_i V^i$, in general different from zero. The gradient of the unit vector $n_\mu$ can also be split into spatial and temporal components as follows
\be
\nabla_{\mu}n_{\nu}=-K_{\mu\nu}-n_{\mu}a_{\nu},
\label{eq:extrinsic}
\ee
where $K_{\mu\nu}$ is the \emph{extrinsic curvature} of the metric (a spatial symmetric tensor) and $a_{\nu}$ is the \emph{acceleration} of the Eulerian observer (a spatial vector too). Finally, it is possible to demonstrate that \citep[e.g.][]{york79}
\be
a_{\nu}=n^{\,\mu}\nabla_{\mu}n_{\nu}=\perp\nabla_{\nu}\ln\alpha,
\ee
another property that will be used later on.

The next step is then to decompose all quantities appearing in the GRMHD equations of Sect.~\ref{sect:cov} into their spatial and temporal components. Hence, we define
\bea
   u^{\,\mu} & = & \Gamma\, n^{\,\mu} + \Gamma\, v^{\,\mu}, 
\label{eq:u} \\ 
T^{\mu\nu} & = & W^{\mu\nu} + S^{\mu}n^{\nu}+ n^{\,\mu}S^{\nu} + Un^{\,\mu}n^{\nu},  
\label{eq:T} \\
F^{\mu\nu} & = & n^{\,\mu}E^{\nu} - E^{\mu}n^{\nu} + \epsilon^{\,\mu\nu\lambda\kappa}B_{\lambda}n_{\kappa}, 
\label{eq:F} \\
F^{*\mu\nu} & = & n^{\,\mu}B^{\nu} - B^{\mu}n^{\nu}  - \epsilon^{\,\mu\nu\lambda\kappa}E_{\lambda}n_{\kappa}, 
\label{eq:F*}
\eea
where all the new vectors and tensors are now spatial and correspond to the familiar three-dimensional quantities as measured by the Eulerian observer. In particular $v^{\,\mu}$ is the usual fluid velocity vector of Lorentz factor $\Gamma=u^{\hat{n}}$, for which
\be
v^i=u^i/\Gamma+\beta^i/\alpha,~~~\Gamma=\alpha u^t=(1-v^2)^{-1/2},
\label{eq:glf}
\ee
where $v^2=v_iv^i$ and we have used the property $u_{\,\mu}u^{\,\mu}=-1$. An alternative quantity, $u^i/u^t=\alpha v^i-\beta^i$, usually referred to as {\em transport velocity}, is sometimes used instead of the Eulerian velocity $v^i$ \citep[see][]{baumgarte03}. The definition in Eq.~(\ref{eq:glf}) agrees with the treatments by \citet{thorne82,sloan85,zhang89} and it is the most appropriate for numerical integration \citep{banyuls97}, since in the $3+1$ formalism $v^i$ is a real three-dimensional vector while $u^i/u^t$ is not. The decomposition of the momentum-energy stress tensor gives the quantities $U=T^{\hat{n}\hat{n}}$, $S^{\mu}=\perp T^{\hat{n}\mu}$, and $W^{\mu\nu}=\perp T^{\mu\nu}$, which are respectively the energy density, the momentum density and the spatial stress tensor of the plasma. Finally, the spatial electromagnetic vectors in Eqs.~(\ref{eq:F}-\ref{eq:F*}) are defined as $E^{\mu}=F^{\hat{n}\mu}$ and $B^{\mu}=F^{*\hat{n}\mu}$, that is, in components
\be
E^i=\alpha F^{ti},~~~~ B^i=\alpha F^{*ti}.
\label{eq:EB}
\ee

\subsection{Derivation of the $3+1$ GRMHD equations}
\label{sect:grmhd}

The set of GRMHD equations in $3+1$ form is derived from that in Sect.~\ref{sect:cov} by applying the space-time decompositions of Eqs.~(\ref{eq:u}-\ref{eq:F*}). Here we are interested in retaining the \emph{conservative} form, as needed by any shock-capturing scheme \citep{font03,shibata05,duez05,anton06}. In this respect, we improve on these works by making use of purely three-dimensional quantities alone, in a way to maintain a close relation to classical MHD as much as possible and to simplify the expression of the source terms. By applying standard covariant differentiation relations, the set of GRMHD equations becomes
\be
(-g)^{-1/2}\partial_\mu [(-g)^{1/2} \rho u^\mu ] = 0,
\ee
\be
(-g)^{-1/2}\partial_\mu [(-g)^{1/2} {T^\mu}_j ] = \textstyle{\frac{1}{2}}T^{\mu\nu}\partial_j g_{\mu\nu},
\ee
\be
(-g)^{-1/2}\partial_\mu [(-g)^{1/2} T^{\mu\nu}n_{\nu} ] = T^{\mu\nu}\nabla_\mu n_{\nu},
\ee
\be
(-g)^{-1/2}\partial_\mu [(-g)^{1/2} F^{*\mu j} ] = 0,
\ee
\be
(-g)^{-1/2}\partial_\mu [(-g)^{1/2} F^{*\mu t} ] = 0,
\ee
where Eqs.~(\ref{eq:mass}), (\ref{eq:momentum}), and (\ref{eq:maxwell2}) have been split into their spatial and temporal components and the symmetry properties of $T^{\mu\nu}$ and $F^{*\mu\nu}$ have been exploited. Eqs.~(\ref{eq:glf}-\ref{eq:EB}) must now be plugged into the above equations to yield equations for the three-dimensional quantities alone. Moreover, it is easy to verify that the source terms on the right hand side are split as
\be
\textstyle{\frac{1}{2}}T^{\mu\nu}\partial_j g_{\mu\nu}=\textstyle{\frac{1}{2}}W^{ik}\partial_j\gamma_{ik}+\alpha^{-1}S_i\partial_j\,\beta^i-U\partial_j\ln\alpha,
\label{eq:source}
\ee
\be
T^{\mu\nu}\nabla_\mu n_{\nu}=-K_{ij}W^{ij}-S^j\partial_j\ln\alpha,
\ee
where the properties of the extrinsic curvature have been used. Notice that only spatial derivatives along $j$ appear in Eq.~(\ref{eq:source}), so that the corresponding flux is a conserved quantity in the stationary case. Finally, it is convenient to introduce the standard boldface notation for (spatial) vectors and to define $\vec{\nabla}=\perp\nabla$ as the three-dimensional covariant derivative operator for the metric $\gamma_{ij}$ (providing the familiar divergence and curl operators), so that the final form of the GRMHD equations is then
\be
\gamma^{-1/2}\partial_t\, (\gamma^{1/2}D) + \vec{\nabla}\cdot (\alpha\vec{v}D-\vec{\beta}D)=0,
\label{eq:cont} 
\ee
\be
\gamma^{-1/2}\partial_t\, (\gamma^{1/2} \vec{S}) + \vec{\nabla}\cdot (\alpha \vec{W}-\vec{\beta}\,\vec{S})= (\vec{\nabla}\vec{\beta})\cdot\vec{S} -U\vec{\nabla}\alpha,
\label{eq:mom} 
\ee
\be
\gamma^{-1/2}\partial_t\, (\gamma^{1/2}U) + \vec{\nabla}\cdot (\alpha \vec{S}-\vec{\beta}U)=\alpha\vec{K}:\vec{W} - \vec{S}\cdot\vec{\nabla}\alpha,
\label{eq:en}
 \ee
\be
\gamma^{-1/2}\partial_t\, (\gamma^{1/2} \vec{B}) + \vec{\nabla}\times (\alpha \vec{E}+\vec{\beta}\times\vec{B})=0,
\label{eq:induct}
\ee
\be
\vec{\nabla}\cdot \vec{B}=0,
\label{eq:divb}
\ee
where $\gamma=\mathrm{det}\{\gamma_{ij}\}$ is the determinant of the spatial metric (not to be confused with the adiabatic index), for which $(-g)^{1/2}=\alpha\gamma^{1/2}$. 

Let us analyze the above system in detail. Eq.~(\ref{eq:cont}) is the continuity equation for $D=\rho\Gamma$, that is the mass density measured by the Eulerian observer. The momentum equation, Eq.~(\ref{eq:mom}), contains the divergence of the tensor $\vec{W}$, leading to source terms present also in MHD and RMHD when curvilinear coordinates are used, whereas the last term with the gradient of the lapse function becomes the standard gravitational force in the Newtonian limit. Eq.~(\ref{eq:en}) is the energy equation, in which the extrinsic curvature must be evolved through Einstein's equations or, for a stationary space-time, it is provided in terms of the covariant derivatives of the shift vector components \citep[e.g.][]{misner73,york79}. Here we write
\be
\alpha\vec{K}:\vec{W}=\textstyle{\frac{1}{2}}W^{ik}\beta^j\partial_j\gamma_{ik}+{W_i}^j\partial_j\,\beta^i,
\label{eq:KW}
\ee
where again the symmetry properties of $W^{ij}$ have been used. Eq.~(\ref{eq:induct}) is the GRMHD extension of the induction equation, written in curl form by exploiting usual vector calculus relations. Note that the (spatial) three-dimensional Levi-Civita tensor density $\epsilon^{\,\mu\nu\lambda}=\epsilon^{\hat{n}\mu\nu\lambda}$, for which $\epsilon^{ijk}=\gamma^{-1/2}[ijk]$ and $\epsilon_{ijk}=\gamma^{1/2}[ijk]$, is implicitly defined in Eq.~(\ref{eq:induct}). Finally, Eq.~(\ref{eq:divb}) is the usual divergence-free condition. Notice that the above treatment is valid in a generic system of curvilinear coordinates, not necessarily under the assumptions of diagonal spatial metric tensor or vanishing expansion factor $\vec{\nabla}\cdot\vec{\beta}$ (e.g. Kerr metric in Boyer-Lindquist coordinates). In the absence of gravity, that is when $\alpha=1$, $\vec{\beta}=0$, $\vec{K}=0$, and $\partial_t\gamma=0$, the above equations reduce to the familiar set of RMHD in curvilinear coordinates.

The expression for the stress tensor, momentum density, and energy density in terms of the fluid and electromagnetic quantities are, from Eqs.~(\ref{eq:u}-\ref{eq:F*}):
\bea
\vec{W} & = & \rho h \Gamma^2\vec{v}\,\vec{v} -\vec{E}\,\vec{E}-\vec{B}\,\vec{B}+[p+\textstyle{\frac{1}{2}}(E^2+B^2)]\,\vec{\gamma},
\label{eq:W} \\
\vec{S} & = & \rho h \Gamma^2\vec{v} + \vec{E}\times\vec{B}, 
\label{eq:S} \\
U & = & \rho h \Gamma^2 - p + \textstyle{\frac{1}{2}}(E^2+B^2), 
\label{eq:U}
\eea
where we have indicated with the symbol $\vec{\gamma}$ the spatial metric tensor of components $\gamma_{ij}$. The matter and electromagnetic field contributions have been expanded by using Eqs.~(\ref{eq:T_matter}-\ref{eq:T_field}) written in terms of scalars and the spatial vectors $\vec{v}$, $\vec{E}$, $\vec{B}$ alone. In the $3+1$ split the Ohm relation for MHD in Eq.~(\ref{eq:mhd}) becomes the usual \emph{freeze-in} condition
\be
\vec{E}=-\vec{v}\times\vec{B},
\label{eq:vxB}
\ee
that allows us to close the set of GRMHD equations. Note that all the above relations, from Eq.~(\ref{eq:W}) to (\ref{eq:vxB}), are exactly the same as in the special relativistic case (though in Paper~II a different formalism was employed). Moreover, the non relativistic limit is found by letting $v^2\ll 1$, $p\ll\rho$, and $E^2\ll B^2\ll\rho$. Thus, by simply changing the definition of $D$, $\vec{W}$, $\vec{S}$, $U$ and by neglecting gravity terms (or reducing them to the Newtonian limit), one has the formal setup of a conservative scheme for classical MHD in generic curvilinear coordinates.

\section{The ECHO scheme}
\label{sect:echo}

The set of conservative GRMHD equations described in Sect.~\ref{sect:grmhd} may be rewritten in a compact way as follows. The five scalar fluid equations are
\be
\partial_t\vec{\mathcal{U}} + \partial_i\vec{\mathcal{F}}^i=\vec{\mathcal{S}},
\label{eq:UFS}
\ee
where the conservative variables and the correspondent fluxes in the $i$ direction are respectively given by
\be
\vec{\mathcal{U}}=\gamma^{1/2}\left[\begin{array}{c}
D \\ S_j \\ U
\end{array}\right],~~~
\vec{\mathcal{F}}^i=\gamma^{1/2}\left[\begin{array}{c}
\alpha v^i D-\beta^i D \\
\alpha W^i_j-\beta^i S_j \\
\alpha S^i-\beta^i U
\end{array}\right],
\label{eq:fluxes}
\ee
and the factors $\gamma^{1/2}$ have been included in the definition of these new quantities. In the case of a \emph{stationary} metric, used in the remainder of this paper for code testing, the source terms become
\be
\vec{\mathcal{S}}=\gamma^{1/2}\left[\begin{array}{c}
0 \\  
\frac{1}{2}\alpha W^{ik}\partial_j\gamma_{ik}+
S_i\partial_j\beta^i-U\partial_j\alpha \\ 
\frac{1}{2}W^{ik}\beta^j\partial_j\gamma_{ik}+{W_i}^j\partial_j\beta^i
-S^j\partial_j\alpha
\end{array}\right],
\ee
in which the extrinsic curvature in the energy equation Eq.~(\ref{eq:en}) has been replaced by the derivatives of the metric according to Eq.~(\ref{eq:KW}). As far as the induction equation is concerned, it is convenient to introduce the new quantities
\bea
\mathcal{B}^i & = & \gamma^{1/2}B^i, \\
\mathcal{E}_i & = & \alpha E_i + \epsilon_{ijk}\beta^jB^k=-[ijk]\mathcal{V}^j\mathcal{B}^k,
\eea
where $\mathcal{V}^j=\alpha v^j-\beta^j$ is the transport velocity. Eq.~(\ref{eq:induct}) may be then rewritten in the form
\be
\partial_t \mathcal{B}^i+[ijk]\partial_j\mathcal{E}_k=0,
\label{eq:induct2}
\ee
and the related non-evolutionary constraint Eq.~(\ref{eq:divb}), expressed in terms of the new variables $\mathcal{B}^i$, simply becomes 
\be
\partial_i {\cal B}^i=0.
\label{eq:divb2}
\ee
Notice that, thanks to our definitions, Eqs.~(\ref{eq:UFS}), (\ref{eq:induct2}), and (\ref{eq:divb2}) retain the same form as in Cartesian coordinates (with external source terms). Eq.~(\ref{eq:induct2}) is the conservation law for $\mathcal{B}^i$, which differs from the form of Eq.~(\ref{eq:UFS}), basically due to the antisymmetric properties of the Faraday and Maxwell tensors. The curl nature of the induction equation and the divergence-free constraint must be maintained in the numerical scheme by employing consistent algorithms.

In the following we describe the numerical procedures employed in our new ECHO code. The scheme is quite general and can be applied to any set of physical laws with evolution equations in the form of Eqs.~(\ref{eq:UFS}-\ref{eq:induct2}), with the additional constraint of Eq.~(\ref{eq:divb2}): physical modules are available for classical MHD, special RMHD, GRMHD, and GRMD (see Sect.~\ref{sect:grmd}). The general recipes for the correct treatment of the divergence-free condition in any shock-capturing MHD-like scheme, regardless of the discretization technique (finite volume or finite difference), accuracy order, interpolation methods, and Riemann solver, have been presented in \citet{londrillo04}. That method was named \emph{Upwind Constrained Transport} (UCT) and here we follow its guidelines. In particular we will adopt the same building blocks already employed in Paper~II, namely finite difference discretization, high order component-wise reconstruction methods (additional algorithms will be proposed here), a two-wave approximate Riemann solver, and multi-stage Runge-Kutta for time integration.

\subsection{Discretization and numerical procedures}
\label{sect:discr}

The starting point is the discretization of the GRMHD equations. Here we assume a finite difference approach and thus we adopt the corresponding version of UCT. This is known to be more convenient than finite volume methods for high order treatments of multi-dimensional problems, since only 1-D reconstruction algorithms are needed \citep[e.g][]{shu97,liu98}. Let $r$ be the order of spatial accuracy requested for the scheme. Given a computational cell of edge sizes $h_i$, the fluid conservative variables $\mathcal{U}_j$ are defined at cell centers $C$ with a \emph{point value} representation, that is $\mathcal{U}_j$ is the numerical approximation, within an accuracy $r$, of the corresponding analytical function. The other conservative variables are the $\mathcal{B}^i$ components, which are here discretized as point values at cell interfaces $S_i^+$, normal to direction $i$. This discretization technique is known as \emph{staggering}, first introduced for Maxwell's equations by \citet{yee66} and later applied to the GRMHD induction equation by \citet{evans88}. In a conservative approach, the spatial differential operators of divergence and curl are translated numerically by making use of the Gauss and Stokes theorems, respectively. Fluid fluxes $\mathcal{F}^i_j$ are to be calculated at cell faces $S_i^+$, while magnetic fluxes $\mathcal{E}_k$ must be calculated at cell edges $L_k^+$, parallel to the direction $k$ \citep[see][]{londrillo04}. The spatially discretized GRMHD equations are then written in the following way
\be
\frac{\mathrm{d}}{\mathrm{d}t}[\mathcal{U}_j]_C+\sum_i\frac{1}{h_i}([\hat{\mathcal{F}}^i_j]_{S_i^+}-[\hat{\mathcal{F}}^i_j]_{S_i^-})=[\mathcal{S}_j]_C,
\label{eq:u_discr}
\ee
\be
\frac{\mathrm{d}}{\mathrm{d}t}[\mathcal{B}^i]_{S_i^+}+\sum_{j,k}[ijk]\frac{1}{h_j}([\hat{\mathcal{E}}_k]_{L_k^+}-[\hat{\mathcal{E}}_k]_{L_k^-})=0,
\label{eq:b_discr}
\ee
known as \emph{semi-discrete} form, since the time derivatives are left analytical. Here the hat indicates high order approximation of the numerical flux function, as it will be described at steps 4 and 8 below, and we have indicated with $\pm$ the opposite faces, or edges, with respect to the direction of derivation. Time evolution is here achieved by means of Runge-Kutta integration schemes. In the same framework, the non-evolutionary solenoidal constraint becomes
\be
\sum_i\frac{1}{h_i}([\hat{\mathcal{B}}^i]_{S_i^+}-[\hat{\mathcal{B}}^i]_{S_i^-})=0.
\label{eq:divb_discr}
\ee

Given the particular discretization of the conservative quantities and of their corresponding numerical fluxes, the procedures required by the UCT strategy may look rather involved, in particular for high order implementations. In the ECHO scheme we have made an effort to simplify them as much as possible, especially as far as the induction equation and the metric terms are concerned. We describe these procedures in the following ten steps.

\begin{enumerate}
\item Given the value of the conservative variables at time $t$, we first interpolate the magnetic field components $\mathcal{B}^i$ from the corresponding staggered locations $S_i^+$ to cell centers $C$, for every direction $i$. For a second order $r=2$ scheme we simply use
\be
[\mathcal{B}^i]_C=\frac{1}{2}([\mathcal{B}^i]_{S^-_i}+[\mathcal{B}^i]_{S^+_i}),
\ee
whereas larger stencils are employed for higher order interpolations (see Sect.~\ref{sect:interp} in the appendix). The set of conservative variables
\be
\vec{\mathcal{W}}=[\vec{\mathcal{U}},\vec{\mathcal{B}}]^T
\ee
is now entirely defined at cell center $C$. From this we can then derive the \emph{primitive} variables $\vec{\mathcal{P}}$, that is any set of physical quantities such that the functions $\vec{\mathcal{U}}=\vec{\mathcal{U}}(\vec{\mathcal{P}})$ and $\vec{\mathcal{F}}^i=\vec{\mathcal{F}}^i(\vec{\mathcal{P}})$ are uniquely defined. Here we use
\be
\vec{\mathcal{P}}=[\rho,\vec{v},p,\vec{B}]^T
\ee
for all MHD-like modules in ECHO. In Sect.~\ref{sect:cons2prim} we describe the inversion routines implemented for this choice of primitive variables.

\item For each direction $i$, say $x$, we reconstruct the point value approximations of the left ($L$) and right ($R$) upwind states of primitive variables, from $C$ to $S_x^+$:
\be
[\mathcal{P}^{L,R}_j]_{S_x^+}=\mathcal{R}^{L,R}_x(\{[\mathcal{P}_j]_C\}),
\ee
where $\mathcal{R}^{L,R}_x$ is the 1-D reconstruction routine, here named REC, applied to a stencil $\{[\mathcal{P}_j]_C\}$ of cell centered values along $x$. The index $j$ runs through all fluid components and the \emph{transverse} magnetic field components. This is because the main assumption in UCT is that the longitudinal $B^x$ component does not present different upwind states at $S_x^+$. At this location one can safely assume ${B^x}^L={B^x}^R=\gamma^{-1/2}\mathcal{B}^x$.

In ECHO different reconstruction routines are implemented. All of them are treated \emph{component-wise}, that is avoiding decomposition into characteristic waves. For schemes with overall $r=2$ accuracy we may use simple TVD-like reconstructions based on limiters (e.g. MM2 for the \emph{MinMod}, MC2 for \emph{Monotonized Centered}). For $r>2$ we have a choice of ENO-like routines: ENO3 for the third-order original ENO method \citep{harten87}, CENO3 for the \emph{Convex}-ENO scheme by \citet{liu98} (see also Paper~I), WENO5 for the \emph{Weighted}-ENO fifth order scheme \citep{jiang96}. Moreover, in the tests of Sect.~\ref{sect:grmhd_tests} and \ref{sect:grmd_tests} we will largely make use of the \emph{Monotonicity Preserving} scheme by \citet{suresh97}, implemented in ECHO as MP5, which is based on interpolation built over a \emph{fixed} 5-point stencil (we recall that adaptive stencils are used in ENO schemes), followed by a filter, basically a combination of limiters to preserve monotonicity near discontinuities. Notice that our reconstruction process is based on upwind, non-oscillatory \emph{interpolation} techniques (thus from point values to point values), while in the numerical literature reconstruction via the primitive function (or equivalently from cell averages to point values) is typically discussed. All interpolation coefficients for high order methods are thus different, and these are calculated in Sect.~\ref{sect:rec} of the appendix.

\item The upwind flux for the fluid part is then derived in terms of the two-state reconstructed primitive variables. In Roe-like schemes \citep{roe81} this task is achieved by a field-by-field spectral decomposition of the local Jacobian $7\times 7$ matrix
\be
\vec{\mathcal{A}}^x=\frac{\partial\vec{\mathcal{F}}^x}{\partial\vec{\mathcal{W}}^x},~~~\vec{\mathcal{W}}^x=[\vec{\mathcal{U}},\mathcal{B}^y,\mathcal{B}^z]^T,
\label{eq:spectral}
\ee
where $\mathcal{B}^x$ acts like a given parameter in this local 1-D system. The eigenvalues of $\vec{\mathcal{A}}^x$, typically calculated at some averaged state, provide the speed of each characteristic wave. Here we use the HLL approximate Riemann solver \citep{harten83} which is based on the knowledge of the two highest (in absolute value) characteristic waves alone. In GRMHD they correspond to the fast magnetosonic waves, see Sect.~\ref{sect:speed}. If $\lambda^x_\pm$ are the requested speeds, calculated at both left and right states, we then define the quantities
\be
a_\pm^x=\mathrm{max}\{0,\pm\lambda^x_\pm (\vec{\mathcal{P}}^L),\pm\lambda^x_\pm (\vec{\mathcal{P}}^R)\}
\label{eq:apm}
\ee
and the HLL upwind fluid flux function is
\be
\mathcal{F}^x_j=\frac{a_+^x {\mathcal{F}^x_j}^L + a_-^x {\mathcal{F}^x_j}^R - a_+^xa_-^x (\mathcal{U}^R_j - \mathcal{U}^L_j )}{a_+^x + a_-^x}
\ee
where all quantities are calculated at $S_x^+$ for each component $j$ and where ${\vec{\mathcal{F}}^x}^{L,R}=\vec{\mathcal{F}}^x(\vec{\mathcal{P}}^{L,R})$, $\vec{\mathcal{U}}^{L,R}=\vec{\mathcal{U}}(\vec{\mathcal{P}}^{L,R})$. At the same location we also calculate the upwind \emph{transverse} transport velocities and we average them as follows
\be
\overline{\mathcal{V}}^j=\frac{a_+^x {\mathcal{V}^j}^L + a_-^x {\mathcal{V}^j}^R}{a_+^x + a_-^x},~~~j=y,z.
\ee
These quantities are saved and will be used at step 6 for the calculation of the electric field needed in the induction equation. The coefficients $a^x_\pm$ are saved too, since they will be needed at step 7 for the magnetic fluxes and at step 10 for the timestep definition. Local Lax-Friedrichs is retrieved as usual when $a_+^x=a_-^x$.

\item The numerical fluid flux function is retrieved by means of an additional high order procedure, named DER, which allows one to obtain a high order approximation from the point value quantities calculated at the same intercell locations:
\be
[\hat{\mathcal{F}}^x_j]_{S_x^+}=\mathcal{D}_x(\{[\mathcal{F}^x_j]_{S_x^+}\}).
\ee
This correction step is necessary to preserve the accuracy in the calculation of spatial partial derivatives for high order schemes, while it can be avoided for low order $r\leq 2$ schemes, for which the DER operator is just an identity. In the tests with $r>2$ presented in Sect.~\ref{sect:grmhd_tests} we use fourth or sixth order fixed-stencil algorithms (see Sect.~\ref{sect:der} in the appendix).

\item The fluid flux functions are recovered for all directions $i$ by repeating steps 2-4 and the spatial operator in Eq.~(\ref{eq:u_discr}) is calculated. The source terms $[\vec{\mathcal{S}}]_C$ are also worked out so that we are ready for the Runge-Kutta time-stepping cycle as far as the fluid part is concerned.

\item The induction equation is treated as follows. Let us concentrate on the magnetic flux $[\hat{\mathcal{E}}_z]_{L_z^+}$, the other components are found with similar strategies. First we need to reconstruct the quantities $\mathcal{V}^x$, $\mathcal{V}^y$, $\mathcal{B}^x$, and $\mathcal{B}^y$ from faces $S_x^+$ and $S_y^+$ to the edge $L_z^+$, to be combined there in a four-state upwind numerical flux \citep{londrillo04}. Exploiting the uniqueness of the numerical representation of $[\mathcal{B}^i]_{S_i^+}$, as discussed at step 2, it is sufficient to reconstruct the following quantities
\be
[{\overline{\mathcal{V}}^x}^{L,R}]_{L_z^+}\!=\mathcal{R}^{L,R}_x(\{[\overline{\mathcal{V}}^x]_{S_y^+}\}),~~[{\mathcal{B}^y}^{L,R}]_{L_z^+}\!=\mathcal{R}^{L,R}_x(\{[\mathcal{B}^y]_{S_y^+}\}),
\ee
\be
[{\overline{\mathcal{V}}^y}^{L,R}]_{L_z^+}\!=\mathcal{R}^{L,R}_y(\{[\overline{\mathcal{V}}^y]_{S_x^+}\}),~~[{\mathcal{B}^x}^{L,R}]_{L_z^+}\!=\mathcal{R}^{L,R}_y(\{[\mathcal{B}^x]_{S_x^+}\}),
\ee
where $\overline{\mathcal{V}}^j$ ($j=x,y$) were saved at step 3.

\item The HLL numerical flux for the magnetic field can be then defined as
\bea
\mathcal{E}_z & = & -\frac{a_+^x {\overline{\mathcal{V}}^x}^L{\mathcal{B}^y}^L + a_-^x {\overline{\mathcal{V}}^x}^R{\mathcal{B}^y}^R - a_+^xa_-^x ({\mathcal{B}^y}^R - {\mathcal{B}^y}^L )}{a_+^x + a_-^x} \nonumber \\
 &  & +\frac{a_+^y {\overline{\mathcal{V}}^y}^L{\mathcal{B}^x}^L + a_-^y {\overline{\mathcal{V}}^y}^R{\mathcal{B}^x}^R - a_+^ya_-^y ({\mathcal{B}^x}^R - {\mathcal{B}^x}^L )}{a_+^y + a_-^y},
\eea
which coincides with the four-state formula presented in \citet{londrillo04}. Note that our flux formula contains upwinding in the two directions $x,y$ and reduces correctly to the expected flux for 1-D cases. 

\item Following the same strategy as in step 4 the DER operation is needed to recover numerical fluxes with appropriate accuracy. Each magnetic flux component actually requires two distinct high order corrections
\be
[\hat{\mathcal{E}}_z]_{L_z^+}=\mathcal{D}_j(\{[\mathcal{E}_z]_{L_z^+}\}),~~~j=x,y
\ee
as Eq.~(\ref{eq:b_discr}) contains both $x$ and $y$ differencing of $\hat{\mathcal{E}}_z$.

\item The spatial derivatives in Eq.~(\ref{eq:b_discr}) are then calculated for each direction and also the induction equation is ready for time integration.

\item Runge-Kutta time-stepping can be finally achieved, and the whole procedure to update the set of conservative variables $\vec{\mathcal{W}}$ must be repeated for each sub-cycle. Here we use for $r\leq 2$ the classical Heun (or improved Euler) second order scheme (RK2), whereas for $r>2$ it is convenient to use correspondingly higher order methods, like those described in \citet{shu88}. In ECHO we have implemented their third order scheme (RK3, see also Paper~I). Like in all explicit schemes, the timestep $\Delta t$ is limited by the CFL (Courant-Friedrichs-Lewy) condition $0<c<1$ (we will always use $c=0.5$ in the tests presented) and is defined as
\be
\Delta t=\frac{c}{\mathrm{max}_i(a^i_M/h_i)},
\ee
where $a^i_M=\mathrm{max}(\{[a^i_+]_{S_i^+}\},\{[a^i_-]_{S_i^+}\})$ are the maximum speeds over the whole domain, for each direction $i$. Gravity contributions to $\Delta t$ are included in the $a^i_M$ definition via the metric terms contained in the GRMHD speeds $\lambda^i_\pm$ (see Sect.~\ref{sect:speed}).

\end{enumerate}

Compared to our previous implementations for classical MHD and RMHD, the ECHO scheme presented here is slightly simpler. First, the DER operator is now based on fixed, symmetric stencils, rather than adaptive like in REC (see the appendix). As far as the induction equation and the related divergence-free constraint are concerned, the use of the magnetic vector potential is avoided and the primary magnetic field (staggered) components for the UCT strategy are now $[\mathcal{B}^i]_{S_i^+}$, rather than $[\hat{\mathcal{B}}^i]_{S_i^+}$ like in \citet{londrillo04}, so that magnetic fields are also easier to initialize. Moreover, it is easy to verify that Eq.~(\ref{eq:divb_discr}) is satisfied algebraically at all times regardless of the value of $r$. This is because, when using Eq.~(\ref{eq:b_discr}) in the time derivative of the solenoidal condition, the electric field components (now with corrections along the \emph{two} orthogonal directions) cancel each other, due to the commutativity of the DER operators applied. Obviously this property holds only for fixed-stencil procedures.

 Finally, notice that the metric terms are needed at cell center (where also their derivatives must be given) and at intercells, but not at cell edges. This is due to our definitions of the $\mathcal{V}^i$ and $\mathcal{B}^i$ components, already containing the metric terms needed for the calculation of the electric field $\mathcal{E}_k$. The components of the metric tensor and their derivatives are here provided analytically. Another option (e.g. when solving Einstein's equations) is to interpolate and derive them, wherever needed, with high order procedures as those described in the appendix.

\subsection{Primitive variables}
\label{sect:cons2prim}

As we have seen in Sect.~\ref{sect:echo}, in step 1 the primitive variables $\vec{\mathcal{P}}$ must be derived from the set of conservative variables $\vec{\mathcal{W}}$ at cell centers. The problem is exactly the same as in special relativistic MHD, that is:
\be
[D,\vec{S},U,\vec{B}] \rightarrow [\rho,\vec{v},p,\vec{B}],
\ee
with $\vec{B}$ acting at the same time as a conservative and primitive variable. Here we basically follow the strategy outlined in Paper~II, see also \citet{noble06} for further discussion and comparison of different techniques. The full system is first reduced to a $2\times 2$ set of nonlinear equations in the variables $x=v^2$ and $y=\rho h\Gamma^2$. Let us rewrite Eqs.~(\ref{eq:S}) and (\ref{eq:U}) using Eq.~(\ref{eq:vxB}) for the electric field, and then calculate $S^2$ and $\vec{S}\cdot\vec{B}$. After some simple algebra, the unknown variables may be found by solving the system $F_1=0$, $F_2=0$, where
\be
F_1(x,y)=(y+B^2)^2x-y^{-2}(\vec{S}\cdot\vec{B})^2(2y+B^2)-S^2,
\label{eq:F1}
\ee
\be
F_2(x,y)=y-p+{\textstyle\frac{1}{2}}(1+x)B^2-{\textstyle\frac{1}{2}}y^{-2}(\vec{S}\cdot\vec{B})^2-U,
\label{eq:F2}
\ee
with $p=p(x,y)$ to be specified according to the EoS employed. Once $x$ and $y$ are found, the required primitive variables are given by the relations
\be
\rho = D(1-x)^{1/2},
\ee
\be
\vec{v} = (y+B^2)^{-1}[\vec{S}+y^{-1}(\vec{S}\cdot\vec{B})\vec{B}],
\ee
\be
p = \frac{\gamma-1}{\gamma}[(1-x)y-D(1-x)^{1/2}],
\label{eq:eos2}
\ee
where the last expression is valid for the ideal gas EoS in Eq.~(\ref{eq:eos}), see \citet{mignone05,ryu06} for other options.

In ECHO the following three inversion methods are implemented.
\begin{enumerate}
\item The roots of Eqs.~(\ref{eq:F1}-\ref{eq:F2}) are found simultaneously via a two-dimensional Newton technique. This system requires a rather accurate initial guess (provided by the quantities found at the previous timestep, at the same grid point) and the inversion of a $2\times 2$ linear system at each iteration.
\item At each iteration, we derive $x=x(y)$ from Eq.~(\ref{eq:F1}) and then we find the root of $f_2(y)\equiv F_2[x(y),y]=0$ by a one-dimensional Newton scheme. This appears to be the most straightforward method, since $x=x(y)$ is just a simple algebraic expression, however in the searching process we must ensure the condition $x<1$ and sometimes several iterations may be required to solve $f_2(y)=0$. 
\item At each iteration, we derive $y=y(x)$ from Eq.~(\ref{eq:F2}) and then we find the root of $f_1(x)\equiv F_1[x,y(x)]=0$ by a one-dimensional Newton scheme. This is a variant of the method suggested in Paper~II and it can only be applied for EoS where $p$ is linear in $y$, as in Eq.~(\ref{eq:eos2}). In this case, the root $y$ is found either simply as a ratio of two terms, if $\vec{S}\cdot\vec{B}=0$, or as the only positive root of the cubic $C(y)$ obtained multiplying Eq.~(\ref{eq:F2}) by $y^2$. This may be achieved either analytically or numerically via a nested Newton scheme. The existence of only one positive root is guaranteed by the following properties: $C(0)<0$, $C^\prime(0)=0$, $C(\pm\infty)=\pm\infty$.
\end{enumerate}
In the tests presented in Sect.~\ref{sect:grmhd_tests} we always use method 3 with the nested Newton procedure to find the root of $C(y)=0$ numerically, since it appears to be rather efficient and robust, especially when applied to a Newton/bisection hybrid method ensuring the search of the solution within given boundaries. In cases of smooth flows where Eq.~(\ref{eq:adiabatic}) replaces the energy equation the inversion algorithm is greatly simplified, since $sD$ is the new conservative variable, hence the pressure $p=s\rho^\gamma$ depends on $x$ alone and we just need to solve the equation $f_1(x)=0$. 

\subsection{Characteristic speeds in GRMHD}
\label{sect:speed}

The spectral properties of the 1-D GRMHD system in Eq.~(\ref{eq:spectral}) are basically the same as for the corresponding system in RMHD. Given the structure of the fluxes it is obvious that, for example, the eigenvalues of the Jacobian $\mathcal{A}^x$ will be of the form
\be
\lambda^x=\alpha {\lambda^\prime}^x - \beta^x,
\label{eq:gr_corr}
\ee
where ${\lambda^\prime}^x$ is the corresponding eigenvalue in special relativistic MHD. Thus, in the $3+1$ approach the gravity terms do not modify substantially the hyperbolic structure of the GRMHD equations. Full descriptions of the spectral decomposition of the 1-D RMHD system in can be found in \citet{anile89}. 

Upwind HLL fluxes, described at step 3, just require the calculation of fast magnetosonic speeds, and this should be accomplished by solving (for each cell and twice for each direction) a quartic polynomial, as already described Paper~II. However, an approximation of these quantities could be also used in Eq.~(\ref{eq:apm}), at a price of slightly higher viscosity. In ECHO we follow the strategy by \citet{gammie03,leismann05}, who realized that, like in classical MHD, an upper bound for fast waves is that corresponding to the degenerate case of normal propagation $k_\mu b^{\,\mu}=0$, where $k_\mu =(-\omega,k_x,0,0)$ is the wave four-vector. The dispersion relation reduces then to
\be
(k_\mu u^{\,\mu})^2=a^2 [(k_\mu k^{\,\mu}) + (k_\mu u^{\,\mu})^2],
\label{eq:disp}
\ee
where the term in square brackets refers to the component of $k_\mu$ normal to $u^{\,\mu}$ and
\be
a^2=c_s^2+c_a^2-c_s^2c_a^2.
\ee
The sound and Alfv\'en speeds are respectively defined as
\be
c_s^2=\frac{\gamma p}{\rho h},~~~c_a^2=\frac{b^2}{\rho h+b^2},
\label{eq:speeds}
\ee
where we have introduced the comoving magnetic four-vector
\be
b^{\,\mu}\equiv F^{*\mu\nu}u_\nu =  \Gamma (\vec{v}\cdot\vec{B}) n^{\,\mu} + B^{\,\mu}/\Gamma + \Gamma (\vec{v}\cdot\vec{B}) v^{\,\mu},
\ee
and the invariant quantity in Eq.~(\ref{eq:speeds}) is
\be
b^2 \equiv b_{\mu}b^{\,\mu} = B^2-E^2 = B^2/\,\Gamma^2+(\vec{v}\cdot\vec{B})^2.
\label{eq:b2}
\ee
In the degenerate case an analytical expression for the two fast magnetosonic characteristic velocities is found by letting ${\lambda^\prime}^x=\omega/k_x$ in Eq.~(\ref{eq:disp}):
\be
\label{eq:magnetosonic}
{\lambda^\prime}^x_\pm\!=\frac{(1\!-\!a^2)v^x\pm\!\sqrt{a^2(1\!-\!v^2)[(1\!-\!v^2a^2)\gamma^{xx}\!-(1\!-\!a^2)(v^x)^2]}}{1-v^2a^2},
\ee
and these upper bounds will be then used also for the general, non-degenerate case. Note that the above relation, when plugged into Eq.~(\ref{eq:gr_corr}), correctly reduces to the $3+1$ GR formula for the hydrodynamical case when $\vec{B}=0$ \citep{banyuls97}.

\subsection{Magnetodynamics}
\label{sect:grmd}

In the present section we summarize the equations of magnetodynamics \citep{komissarov02,komissarov04a} and we discuss the few modifications implemented in ECHO for the corresponding GRMD module. The recipes by \citet{mckinney06a}, which allow one to use the same framework of a GRMHD scheme and simply neglect the matter contribution, are here followed. In GRMD the fluid quantities disappear and the electric field $\vec{E}$ should replace them as primary variable, together with $\vec{B}$. The equations to use should be then the two Maxwell equations Eqs.~(\ref{eq:maxwell1}-\ref{eq:maxwell2}), like in electrodynamics. However, here we replace Eq.~(\ref{eq:maxwell1}) with the electromagnetic momentum-energy conservation law. Thus, by setting $T^{\mu\nu}\simeq T^{\mu\nu}_f\gg T^{\mu\nu}_m$ in Eqs.~(\ref{eq:momentum}) and (\ref{eq:lorentz}) in the limit of negligible plasma inertia and thermal contribution, we find
\be
\nabla_{\mu}T^{\mu\nu}=J_{\mu}F^{\mu\nu}=0.
\ee 
This \emph{force-free} situation is actually common to vacuum electrodynamics as well. However, in a highly conducting plasma we assume that there is a frame where the electric field vanishes, due to the presence of freely moving charges always able to screen it efficiently, just like in the GRMHD approximation. This is the reason why magnetodynamics is commonly known as \emph{degenerate} force-free electrodynamics. If the electromagnetic fields are decomposed according to the Eulerian observer in the $3+1$ approach of Sect.~\ref{sect:3+1}, the condition for the existence of a frame where the electric field vanishes is replaced by the two invariant conditions
\be
B^2-E^2\geq 0,~~~\vec{E}\cdot\vec{B}=0,
\label{eq:constr}
\ee
which are valid in GRMHD too thanks to ideal Ohm's law Eq.~(\ref{eq:vxB}). If we still indicate with $u^{\,\mu}$ the unit time-like four-velocity of this frame, and $\vec{v}$ is the associated three-velocity defined in Eq.~(\ref{eq:glf}), the usual ideal MHD condition is unchanged and the two constraints in Eq.~(\ref{eq:constr}) are automatically satisfied. In order to close the GRMD system, we thus need to express this unknown velocity in terms of the electromagnetic quantities alone. The required $\vec{v}$ turns out to be the \emph{drift} speed of magnetic fieldlines
\be
\vec{v}=\frac{\vec{E}\times\vec{B}}{B^2}.
\label{eq:drift}
\ee
All the (G)RMHD definitions in Eqs.~(\ref{eq:U}) to (\ref{eq:vxB}) are still valid if one neglects matter contribution, in particular $\vec{S}=\vec{E}\times\vec{B}$. Notice that due to Eqs.~(\ref{eq:vxB}) and (\ref{eq:drift}) the three spatial vectors $\vec{E}$, $\vec{B}$, and $\vec{v}$ are all mutually orthogonal in GRMD. When the three-velocity in Eq.~(\ref{eq:drift}) is used, the equations for GRMHD remain unchanged too. However, the continuity equation Eq.~(\ref{eq:cont}) is now useless, while the energy equation Eq.~(\ref{eq:en}) is redundant and may be used as an additional check. Notice that, in particular, the treatment of the metric terms and of their derivatives in the source part remains exactly the same as in GRMHD.

From a computational point of view, the set of GRMD in conservative form is easy to treat. The characteristic speeds are two Alfv\'en waves and two magnetosonic waves, moving at the speed of light. Thus, the expression needed for the simplified Riemann solver employed in ECHO (along the $x$ direction) is derived from Eqs.~(\ref{eq:gr_corr}) and (\ref{eq:magnetosonic}) by setting $a=1$, that is
\be
\lambda^x_\pm=\pm \alpha\sqrt{\gamma^{xx}}-\beta^x.
\ee
Furthermore, the inversion from conservative to primitive variables is also greatly simplified. The magnetic field still enters both as a conservative and primitive variable, hence we need to derive the drift velocity $\vec{v}$ for given $\vec{S}$ and $\vec{B}$. The expression employed in ECHO is
\be
\vec{v}=\frac{1}{B^2}\left[\vec{S}-\frac{(\vec{S}\cdot\vec{B})}{B^2}\vec{B}\right],
\ee
where the second term takes into account the possible numerical errors leading to an initial non-vanishing $\vec{S}\cdot\vec{B}$. Notice that the above formula is equivalent to first derive the electric field as $\vec{E}=-\vec{S}\times\vec{B}/B^2$ and then use Eq.~(\ref{eq:drift}). In this way, our code preserves the constraint $\vec{E}\cdot\vec{B}=0$ within machine accuracy. 

\section{GRMHD numerical tests}
\label{sect:grmhd_tests}

In order to test our numerical scheme ECHO, several aspects need to be checked. First we want to verify that in spite of the UCT algorithm, based on staggered representation of the magnetic field components, the overall scheme is able to preserve the nominal high order accuracy of the reconstruction and interpolation routines employed. Hence we propose a new test based on the propagation of Alfv\'en waves (in flat space-time), which are \emph{smooth} solutions of the equations and thus suitable for such kind of problems. However, to better compare ECHO's performances against other existing GRMHD codes, we will employ ECHO at second order in most of the other numerical test problems. Thus, even if higher than second order reconstruction algorithms will be used, in order to sharpen discontinuities and reduce numerical diffusion (in particular MP5), all additional corrections to achieve an effective higher order of spatial accuracy will be sometimes disabled and RK2 will be used for time stepping in these cases. We will see that the resulting second order scheme (much simpler to be implemented) is a good compromise between efficiency, accuracy, and robustness. The other numerical tests considered here are: 1-D and 2-D problems to check the code shock-capturing properties (a shock tube and the cylindrical blast wave); 1-D accretion onto black holes, in Schwarzschild and Kerr metrics, to verify ECHO's high order properties in curved space-times too; stability of a thick disk (with constant angular momentum and with a toroidal magnetic field) around a Kerr black hole as a test in 2-D GRMHD. All the problems discussed here will involve the presence of substantial magnetic fields with plasma beta (the ratio of thermal to magnetic pressure) of order of unity or lower.

If not differently stated, in all our numerical tests we will use a Courant number of 0.5, a $\gamma$-law EoS with $\gamma=4/3$, and we will solve the equation for the total energy density $U$. Grid spacing will always be constant (though non-uniform grids are permitted in ECHO), so the number of points is enough to specify the grid in each direction (a single grid point is assigned to the ignorable coordinates).

\subsection{Large amplitude CP Alfv\'en wave}

The first test we propose here is a novel one, not previously employed in other works on numerical relativistic MHD to our knowledge. It involves the propagation of \emph{large amplitude} circularly polarized (CP) Alfv\'en waves along a uniform background field $\vec{B}_0$ in a numerical domain, 1-D or 2-D, with periodic boundary conditions. Since the propagating wave is an exact solution, as we will see below, the test is very useful to check the accuracy (both spatial and temporal) and spectral resolution properties of a numerical scheme. This is achieved by measuring the errors in the solution after one or more periods compared to the initial conditions. Such test was first proposed in our Paper~II in the case of small amplitudes, where the solution was only an approximate one. Here we show how to extend the exact solution valid in the non-relativistic case to the most general case of large amplitudes in (special) relativistic MHD. For the general properties of Alfv\'enic modes in RMHD see \citet{anile89} and \citet{komissarov97}, for other (but less straightforward) numerical tests involving a different kind of Alfv\'enic exact solutions see \citet{komissarov99} and \citet{duez05}.

\begin{table}
\centering
\begin{tabular}{lrcccc}
\hline\hline
& & \multicolumn{2}{c}{1-D} & \multicolumn{2}{c}{2-D} \\
Method & $N$ & $L_1$ error & $L_1$ order & $L_1$ error & $L_1$ order \\
\hline
MC2   &   8 & 1.58e-1 &  --  & 1.81e-1 &  --  \\
      &  16 & 3.63e-2 & 2.12 & 4.60e-2 & 1.98 \\
      &  32 & 7.14e-3 & 2.34 & 8.23e-3 & 2.48 \\
      &  64 & 1.55e-3 & 2.20 & 1.71e-3 & 2.27 \\
      & 128 & 3.69e-4 & 2.07 & 4.01e-4 & 2.09 \\
      & 256 & 8.98e-5 & 2.04 & 9.76e-5 & 2.04 \\
      & 512 & 2.21e-5 & 2.02 &    --   &  --  \\
CENO3 &   8 & 8.25e-2 &  --  & 1.07e-1 &  --  \\
      &  16 & 1.25e-2 & 2.72 & 1.68e-2 & 2.67 \\
      &  32 & 1.65e-3 & 2.92 & 2.21e-3 & 2.92 \\
      &  64 & 2.09e-4 & 2.98 & 2.80e-4 & 2.98 \\
      & 128 & 2.62e-5 & 3.00 & 3.50e-5 & 3.00 \\
      & 256 & 3.28e-6 & 3.00 & 4.38e-6 & 3.00 \\
      & 512 & 4.10e-7 & 3.00 &    --   &  --  \\
WENO5 &   8 & 3.91e-2 &  --  & 4.76e-2 &  --  \\
      &  16 & 2.35e-3 & 4.06 & 3.14e-3 & 3.92 \\
      &  32 & 8.73e-5 & 4.75 & 1.16e-4 & 4.76 \\
      &  64 & 2.82e-6 & 4.95 & 3.76e-6 & 4.95 \\
      & 128 & 8.96e-8 & 4.98 & 1.19e-7 & 4.98 \\
      & 256 & 2.79e-9 & 5.01 & 3.71e-9 & 5.00 \\
      & 512 & 8.53e-11& 5.03 &    --   &  --  \\
MP5   &   8 & 1.05e-2 &  --  & 1.37e-2 &  --  \\
      &  16 & 3.71e-4 & 4.82 & 4.98e-4 & 4.78 \\
      &  32 & 1.20e-5 & 4.95 & 1.16e-5 & 4.95 \\
      &  64 & 3.82e-7 & 4.97 & 5.08e-7 & 4.99 \\
      & 128 & 1.20e-8 & 4.99 & 1.59e-8 & 5.00 \\
      & 256 & 3.75e-10& 5.00 & 4.98e-10& 5.00 \\
      & 512 & 1.21e-11& 4.95 &    --   &  --  \\
\hline
\end{tabular}
\caption{Accuracy for the CP Alfv\'en wave test. The $L_1$ errors and orders are shown for various methods as a function of the number of grid points, both in 1-D and 2-D. Notice that only when the error becomes lower than $\sim 10^{-10}$ (the value of the tolerance in the inversion from conservative to primitive variables) discrepancies from the nominal order start to appear.}
\label{tab:alfven}
\end{table}

\begin{figure*}
\centering
\resizebox{\hsize}{!}{
\includegraphics{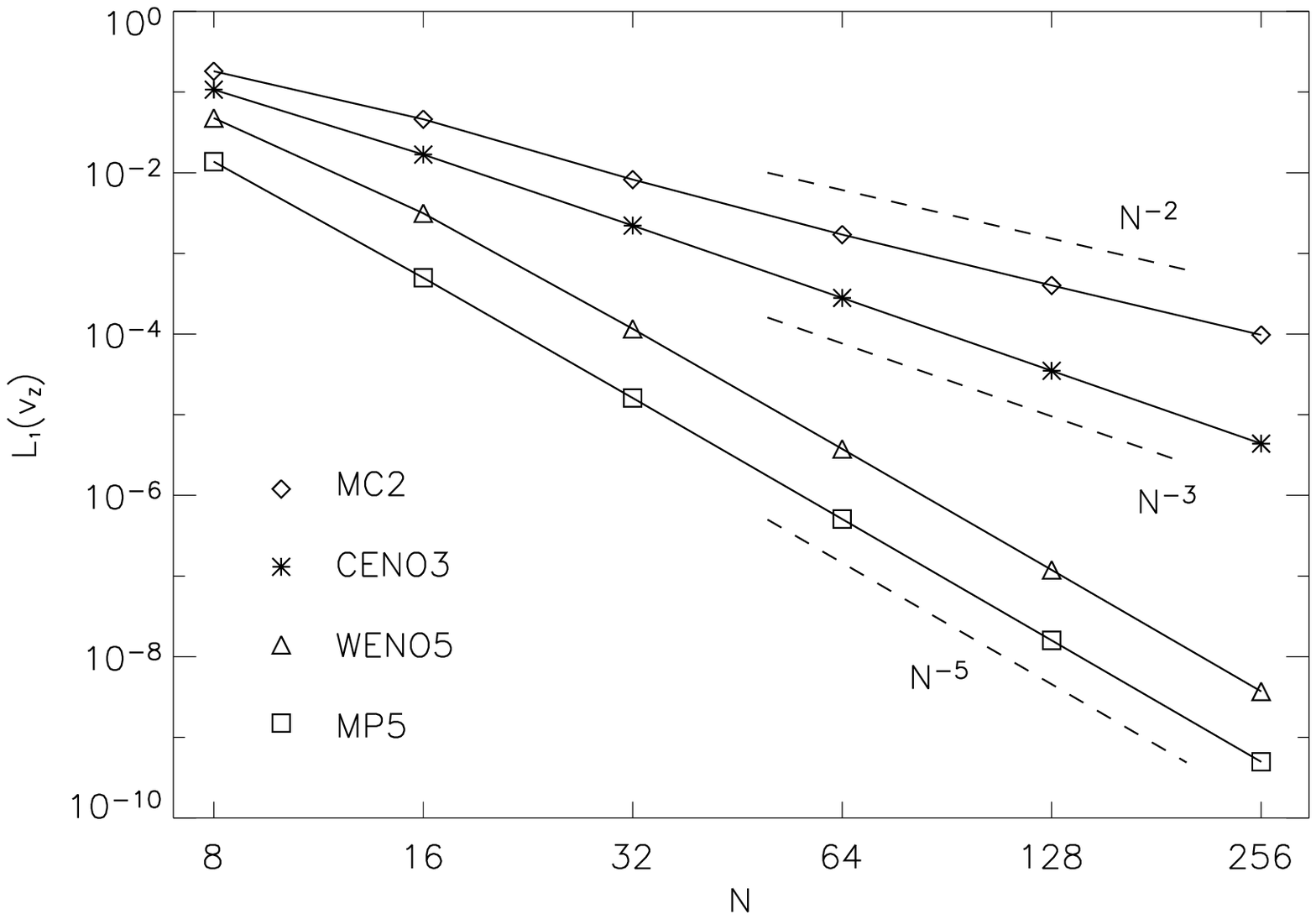}
\includegraphics{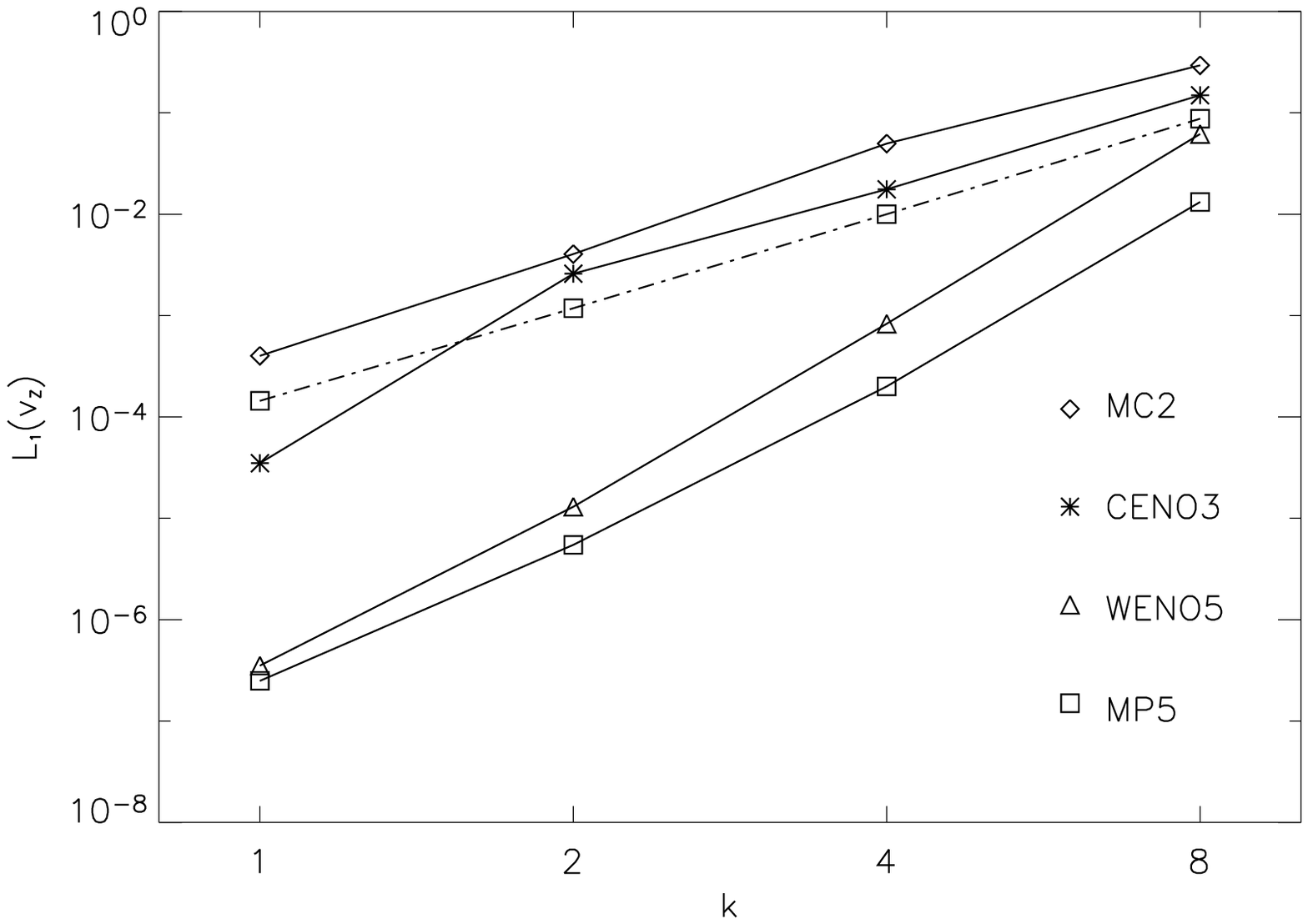}
}
\caption{
The large amplitude CP Alfv\'en wave test in the 2-D case (propagation along the diagonal). The $L_1$ errors for the $v_z$ velocity component, obtained by comparing the solution at the final time $t$ with respect to the initial conditions, for different interpolation schemes. In the left panel we show the dependence on the number of grid points $N$ (for a fixed wave number $k=1$), whereas in the right panel we show the dependence on $k$ for a fixed resolution of $128\times 128$. The dashed-dotted line in this second plot refers to the run with MP5 at overall second order, and it roughly corresponds to a straight line with $L_1\sim k^3$.
}
\label{fig:alfven}
\end{figure*}

Let us consider a CP Alfv\'en wave of normalized amplitude $\eta$. In classical MHD the variable quantities are the transverse components of $\vec{B}$ and $\vec{v}$, which are parallel to each other with vector tips describing circles in the plane normal to $\vec{B}_0$. Whatever the wave amplitude, these are the only fluctuating fields and the background quantities are not affected by the wave (in particular $\rho$ and $p$, since the wave is incompressible). In the RMHD case, let us look for an exact solution with the same properties. The transverse components of $\vec{B}$ are written
\be
B_y=\eta B_0\cos[k(x-v_A t)],~~~~B_z=\eta B_0\sin[k(x-v_A t)],
\label{eq:CP}
\ee
where we have assumed $B_x=B_0$, $v_A$ is the (still unknown) Alfv\'en speed, and $k$ is the wave vector. Since the induction equation remains exactly the same as in the non-relativistic case, we still take the velocity components in the form (let us take $v_x=0$ for simplicity): 
\be
v_y=-v_A B_y/B_0,~~~v_z=-v_A B_z/B_0,
\ee
as in the classical MHD, where in that case $v_A=B_0/\rho^{1/2}$ whatever the wave amplitude $\eta$ (the minus sign gives propagation in the positive $x$-direction). We will now see that in the relativistic case this value is different, basically due to the contribution of the kinetic and electromagnetic energies to the inertia of the plasma and to the presence of no longer negligible electric forces in the momentum equation. 

The electric field is derived from Eq.~(\ref{eq:vxB}), so $E_y=-v_zB_x=v_AB_z$, $E_z=v_yB_x=-v_AB_y$, $E_x=-v_yB_z+v_zB_y=0$. Notice also that the quantities $v^2=\eta^2 v_A^2$, $B^2=B_0^2(1+\eta^2)$, and $E^2=\eta^2 v_A^2 B_0^2$ are constant, as well as $\rho$ and $p$ (hence $h$ too). It is easy to show that the transverse components of the momentum equation yield the condition
\be
\label{eq:v_A}
[\rho h+(1+\eta^2-\eta^2v_A^2)B_0^2]v_A^2=B_0^2,
\ee
where in square brackets we have the total enthalpy $\rho h+B^2-E^2$, which depends on $v_A$ itself. Eq.~(\ref{eq:v_A}) is a second order algebraic equation for $v_A^2$, where in order to preserve the condition $v_A^2<1$ the smaller solution must be chosen. Rearranging the terms we finally find
\be
v_A^2=\frac{B_0^2}{\rho h\!+\!B_0^2(1\!+\!\eta^2)}\left[\frac{1}{2}\left(1\!+\sqrt{1\!-\left(\frac{2\eta B_0^2}{\rho h\!+\!B_0^2(1\!+\!\eta^2)}\right)^2}\,\right)\right]^{-1}\!\!.
\ee
Notice that in the small amplitude limit $\eta\ll 1$ we retrieve the familiar expression $v_A^2=B_0^2/(\rho h+B_0^2)$ used in Paper~II. When we further have $h\ll 1$ and $B_0^2\ll\rho$ the classical MHD limit $v_A^2=B_0/\rho$ is found, as expected.

From a numerical point of view, we test the accuracy of our scheme by measuring the errors on one of the transverse quantities, say $v_z$, at time $t=T=L/v_A$ (one period), compared to the initial condition in Eq.~(\ref{eq:CP}) at $t=0$. For the 1-D case we take a periodical numerical domain along $x$ of length $L=2\pi$, while in the 2-D case we rotate the initial conditions in the $(x,y)$ plane so to have propagation along the diagonal of a bi-periodical $[0,2\pi]^2$ domain. As discussed in Paper~II, now two complete spatial periods are contained along the diagonal of length $L=2\pi\sqrt{2}$, so we can take $t=T/2$ as final time. With the above choices the wave vector $k$ coincides with the wave number, hence it corresponds to the (integer) number of spatial periods present in the numerical domain. For this test we normalize our physical quantities by assuming $\rho=p=B_0=\eta=1$.

In Table~(\ref{tab:alfven}) we show the errors and convergence orders in the $L_1$ norm (the absolute error averaged over the whole computational domain) for the test with $k=1$ at various resolutions. This is done for both the 1-D and 2-D cases, and for different reconstruction schemes. The errors for the 2-D case are also plotted in the left panel of Fig.~(\ref{fig:alfven}). Note that the nominal order of accuracy is achieved already at small numbers $N$ of grid points, which means that basically the reconstruction routines employed always use the full stencil at their disposal, as expected for smooth solutions, without dropping to lower orders at wave extrema. In order to achieve third order convergence the RK3 time stepping algorithm has been employed for CENO3, while to be able to reach an overall fifth order in time and space for WENO5 and MP5 the RK3 routine has been used with $\Delta t\propto N^{-5/3}$, so that the accuracy in time becomes of order $O(\Delta t^3)=O(N^{-5})=O(\Delta x^5)$, i.e. the same of the spatial one, as needed in this kind of tests \citep[e.g.][]{jiang96}. Here the best performing schemes are obviously those with higher nominal orders (that for smooth solutions), thus WENO5 and MP5 in our case. In spite of the same fifth order of accuracy, MP5 return smaller errors, up to almost a factor 10 at high resolution. This demonstrates that the limiting conditions in MP5 never apply for this test and the optimal stencil is always used, whereas the weights in the WENO5 routine do not precisely match to provide the corresponding optimal stencil. 

Finally, we test the spectral resolution of our schemes by running the same problem at various wave numbers $k$, from 1 to 8, at a fixed resolution of $128\times 128$ (in the 2-D case). The $L_1$ error now increases with $k$, as expected (an increasing $k$ basically means a decreasing resolution), and the dependence is stronger for increasing orders $r$. Indeed, higher order schemes (here without the correction to the timestep, thus with a third order temporal accuracy at most) are able to reproduce reasonably the analytical solution even at the smallest wavelengths, where second order schemes give poor results. A good compromise between efficiency and accuracy is MP5 with RK2 time-stepping ($3/2$ times faster than RK3) and without higher order corrections (with overall $r=2$ second order accuracy), which appears to behave better than CENO3 with RK3 at small wavelengths.

\subsection{Shock tube with gauge effects}

\begin{figure*}
\centering
\includegraphics[height=10.5cm]{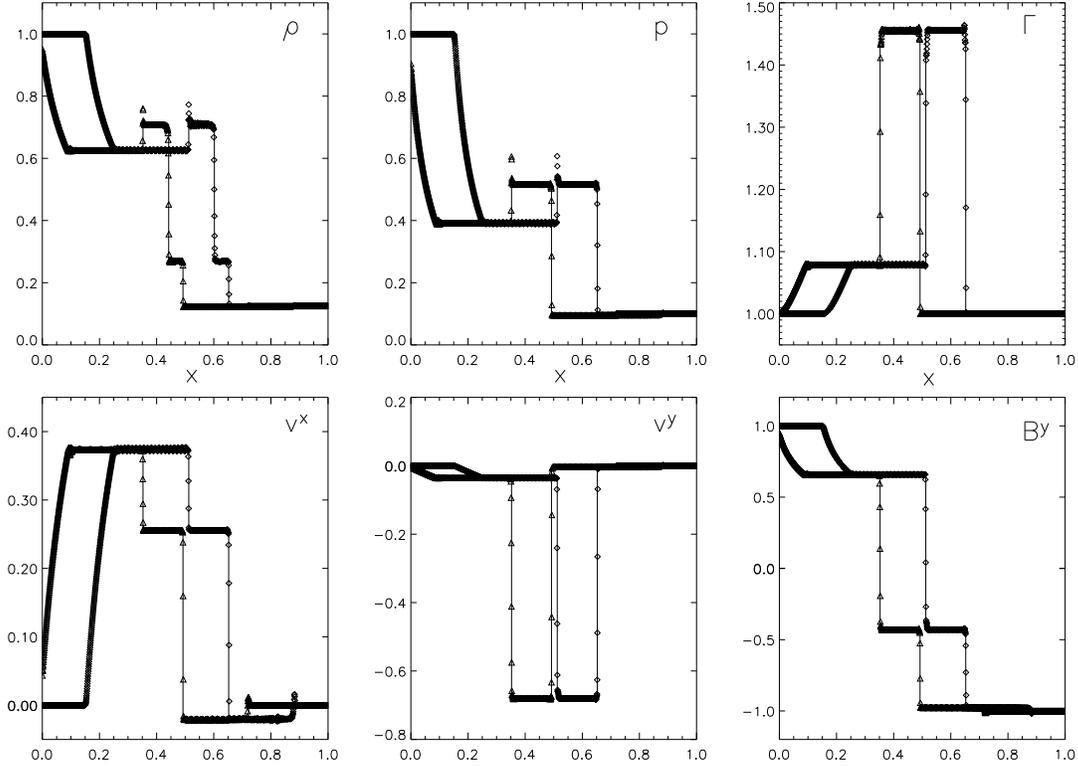}
\caption{
The relativistic Brio \& Wu shock tube modified to allow for gauge effects. The solution on the right hand side refers for a run with $\alpha=2$, $\beta^x=0$, and $t=0.2$ (diamonds), whereas that on the left hand side to a run with $\alpha=1$, $\beta^x=0.4$, and $t=0.16$ (triangles). The numerical solutions are over-plotted to the results obtained with an exact Riemann solver (solid line).Both tests are computed with MP5 (no DER and RK2) and $N=1600$ grid points.
}
\label{fig:st}
\end{figure*}

\begin{figure}
\centering
\resizebox{\hsize}{!}{\includegraphics{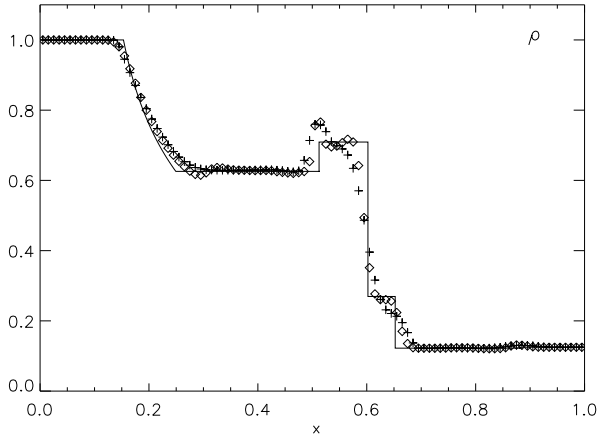}}
\caption{
Comparison of different schemes in the relativistic Brio \& Wu shock tube test. Only $N=100$ grid points are used and the density profile is shown for $t=0.4$. Results obtained with REC based on MP5 (diamonds) appear less smearing than those obtained with MC2 (pluses), at the price of some oscillations.
}
\label{fig:st_comp}
\end{figure}

\begin{figure*}
\centering
\includegraphics[height=6.8cm,width=6cm]{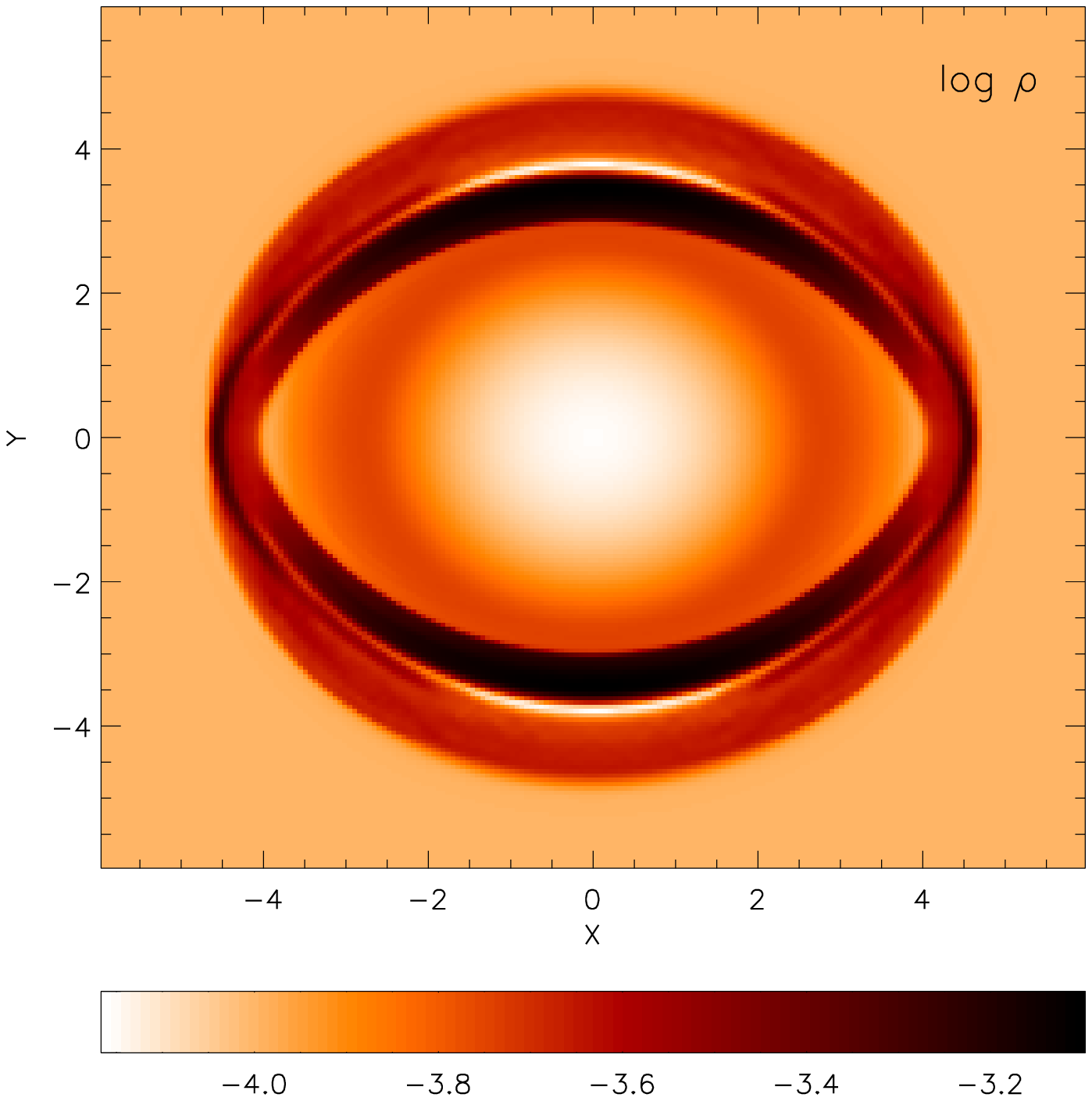}
\includegraphics[height=6.8cm,width=6cm]{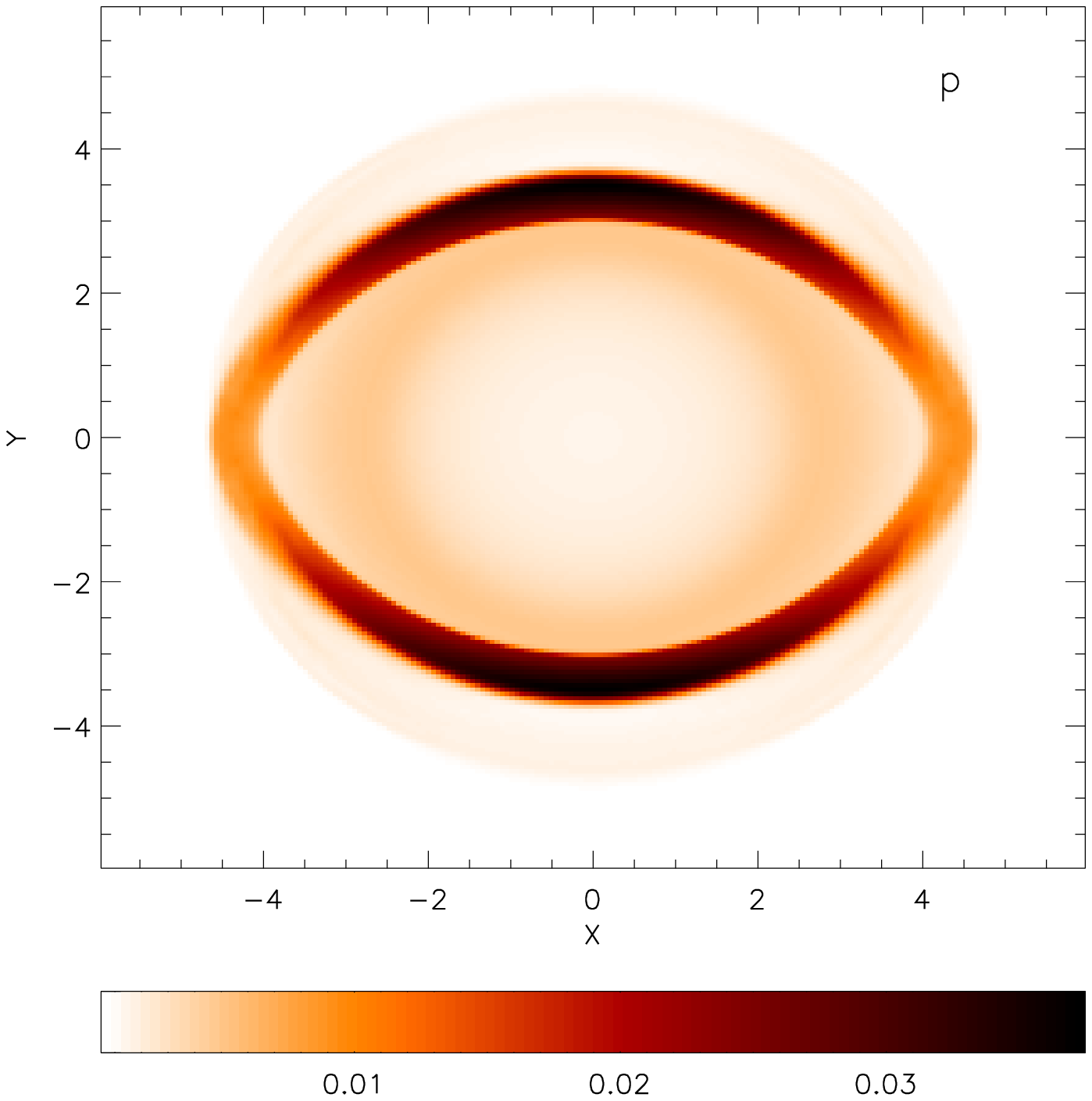}
\includegraphics[height=6.4cm,width=6cm]{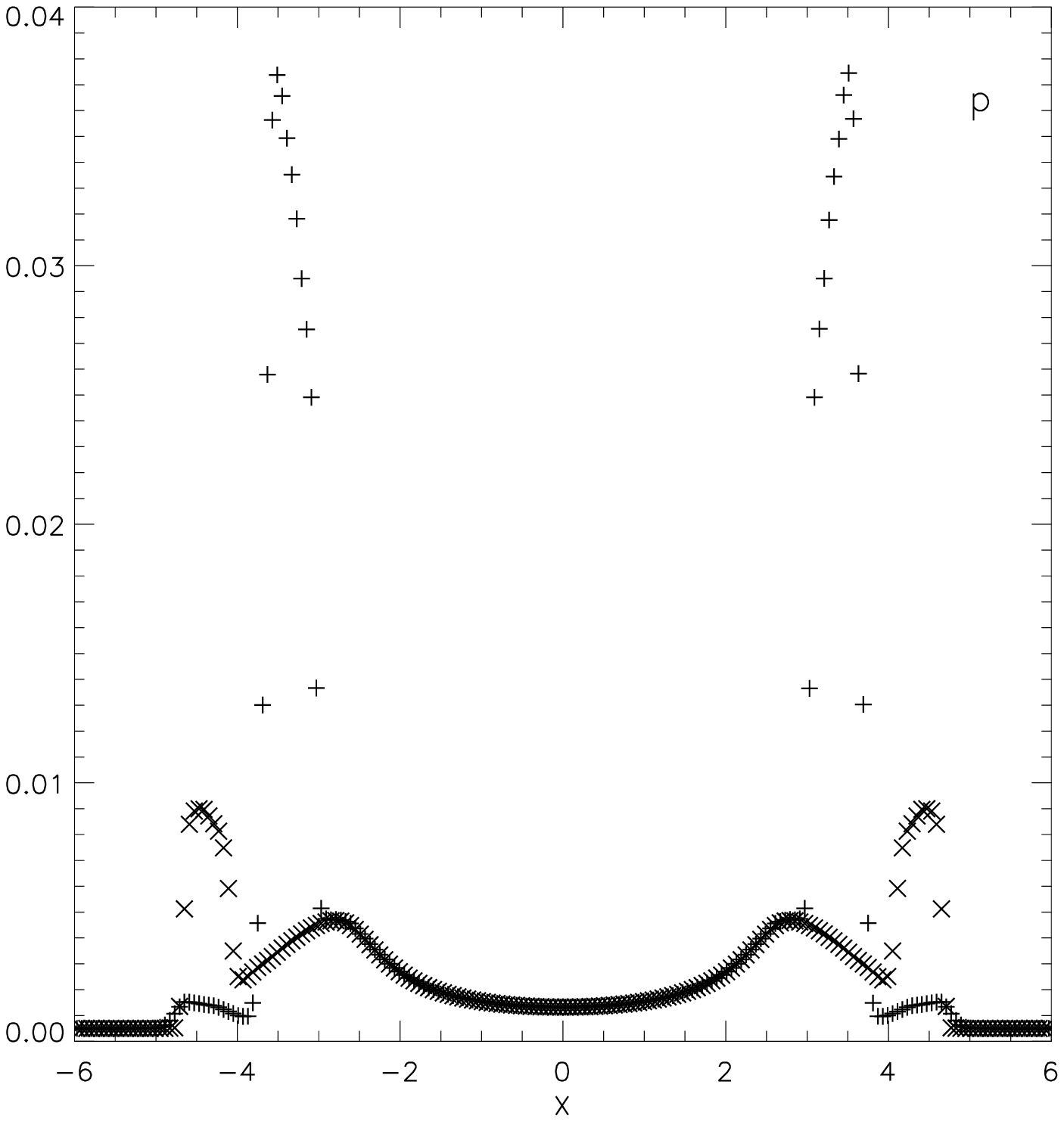}
\includegraphics[height=6.8cm,width=6cm]{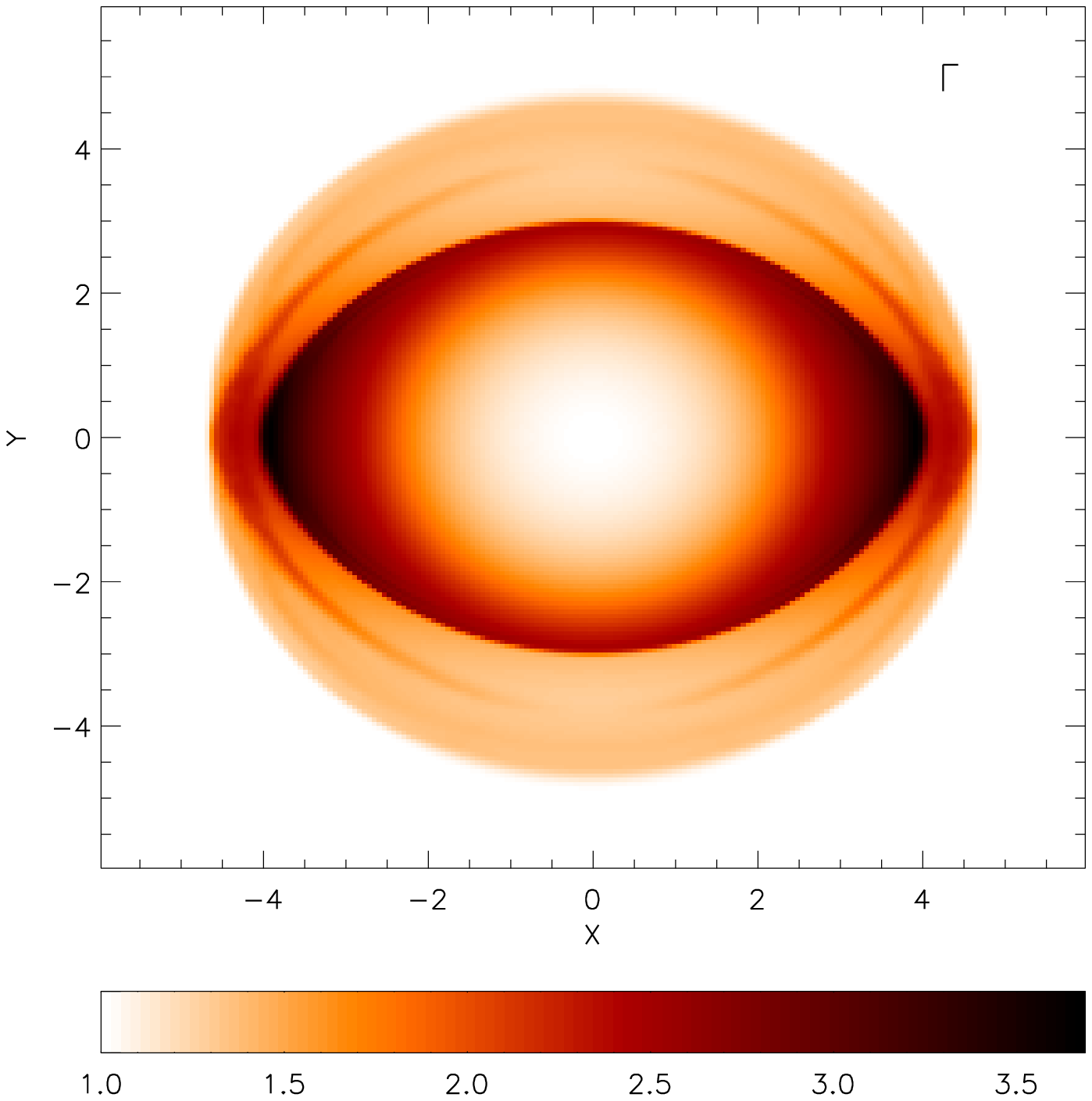}
\includegraphics[height=6.8cm,width=6cm]{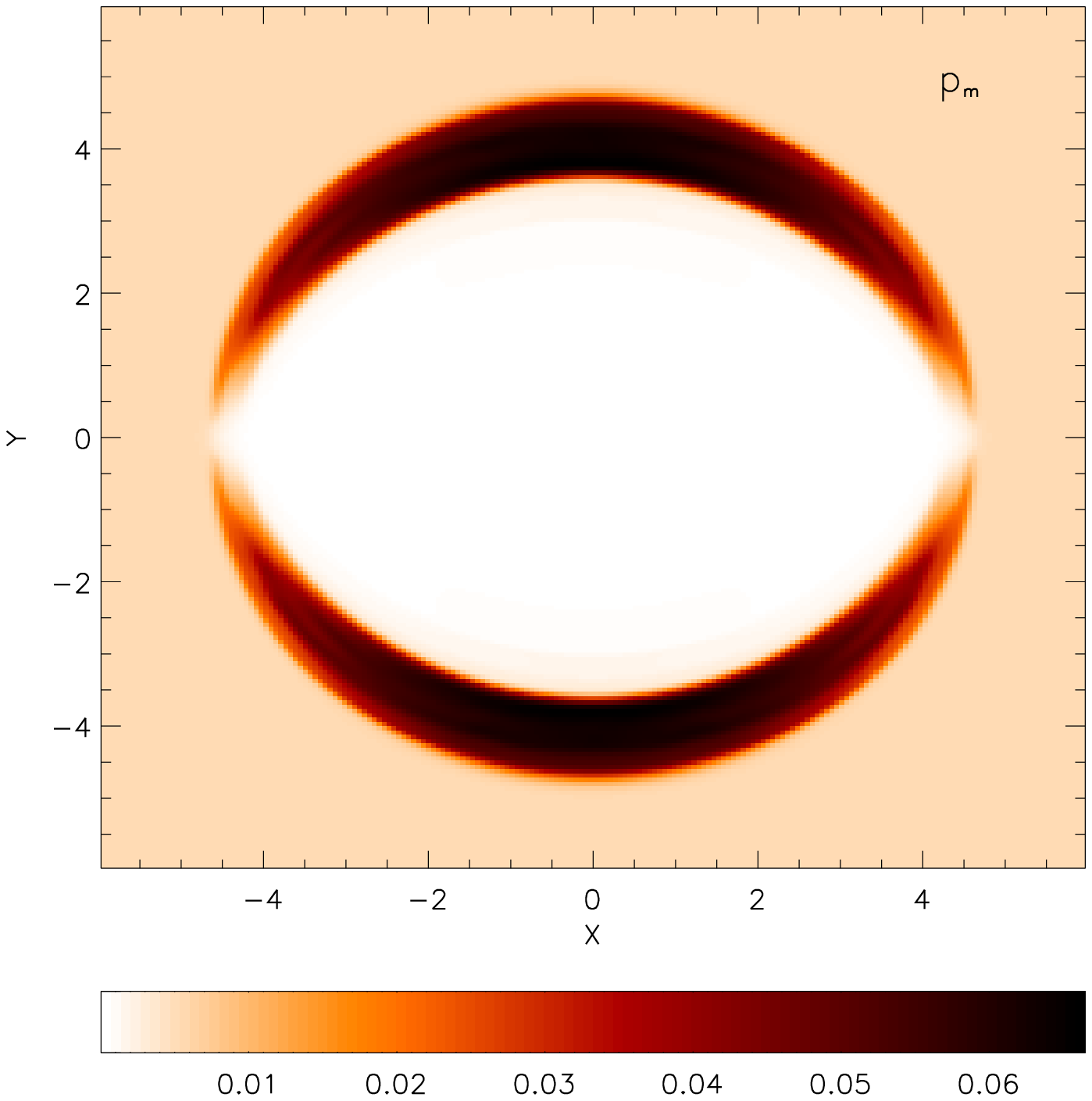}
\includegraphics[height=6.4cm,width=6cm]{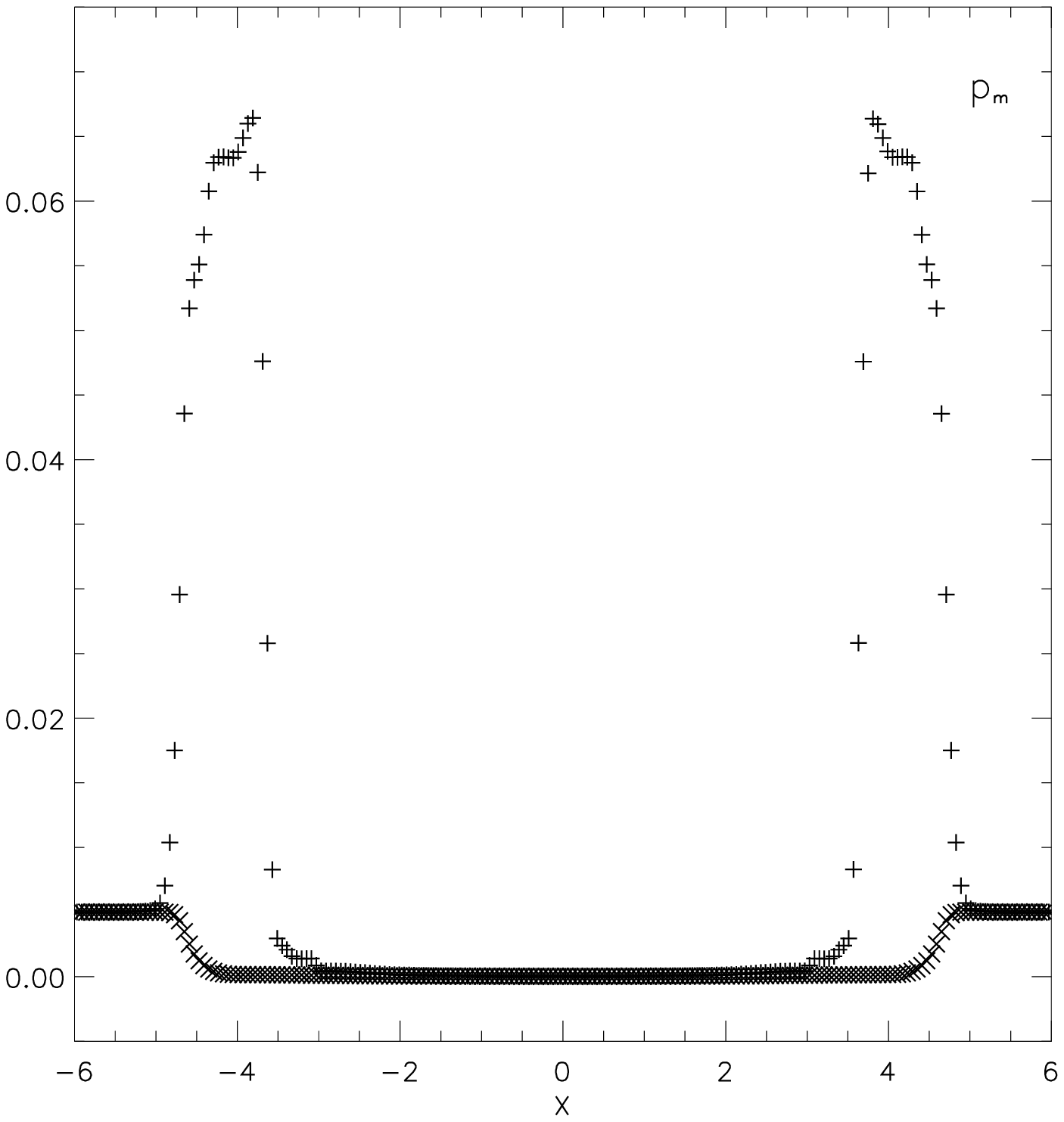}
\caption{
The magnetized cylindrical blast wave test at $t=4.0$. 2-D maps of density $\rho$, thermal pressure $p$ , Lorentz factor $\Gamma$, and magnetic pressure $p_\mathrm{m}=B^2/2$ are shown on the four panels on the left hand side. On the two plots on the right we show cuts through the center of the domain for thermal (upper panel) and magnetic (lower panel) pressure. Horizonthal cuts are indicated with crosses, while pluses are used for vertical cuts.
}
\label{fig:blast2d}
\end{figure*}

Shock tubes are excellent tests to monitor the shock-capturing properties of a numerical scheme. Until recently, however, an exact solver for (special) relativistic MHD Riemann problems was still missing, so that comparison was simply made by running the code at different resolutions and relying on the convergence properties of the conservative numerical scheme employed. Now the situation has changed and we can test our numerical solutions against the exact Riemann solver for RMHD by \citet{giacomazzo06}, kindly provided by the authors. Since RMHD shock tubes have been extensively presented in Paper~II, here just an example will be given, namely the relativistic version of the Brio \& Wu test \citep{brio88} by \citet{vanputten93} and \citet{balsara01}. The initial conditions are 
\be
(\rho,p,B^x,B^y)=\left\{ \begin{array}{rrrrrl} 
( &   1.0, & 1.0, & 0.5, &  1.0), & x<0.5 \\
( & 0.125, & 0.1, & 0.5, & -1.0), & x>0.5,
\end{array}\right.
\ee
while the other quantities are set to zero. A $\gamma-$law EoS with $\gamma=2$ is used, and the final time is $t=0.4$. Following \citet{anton06}, instead of showing the standard RMHD results, we turn here the test in a sort of GRMHD problem by choosing different gauges while preserving a flat metric. In Fig.~(\ref{fig:st}) we show the numerical results obtained by using $\alpha=2.0$ (diamonds), compared with the exact solution plotted for $t/\alpha=0.2$, and those obtained with $\beta^x=0.4$ (triangles), compared with the exact solution shifted by $\delta x=\beta^x t=0.16$. For both runs MP5 is used (no DER and RK2), and $N=1600$ grid points are employed.

The first thing to notice is that all the usual structures arising from the breakout of the initial discontinuity (left-going fast rarefaction wave, left-going slow compound wave, contact discontinuity, right-going slow shock, right-going fast rarefaction wave) are well reproduced in both cases, so the chosen gauges work as expected. In particular, note the presence near the initial discontinuity position $x=0.5$ (in the $\alpha=2.0$ test) of the so-called \emph{compound} wave, here appearing as a discontinuity. This is the combination of an intermediate shock and a rarefaction wave, a feature sometimes encountered in coplanar problems due to the non-strict hyperbolicity of MHD. Given its nature, it cannot be found by exact Riemann solvers and the physical acceptability itself as solution of the \emph{ideal} MHD equations is still debated \citep{barmin96,myong98,torrilhon04}. On the other hand, this feature is invariably found by means of any numerical scheme, where some sort of dissipation, either physical or numerical, is always present. As far as the reconstruction algorithm is concerned, we can see that MP5 gives sharp profiles at all discontinuities, which are captured within $5-10$ grid points.

In Fig.~(\ref{fig:st_comp}) we show a comparison of the reconstruction (REC) performances of the scheme for the same test, now with the original settings ($\alpha=1$, $\beta=0$ and $t=0.4$). Here we use low resolution runs ($N=100$ grid points) to better appreciate the differences. The two reconstructions are MP5 and MC2, both at an overall second order in space (non DER) and time (RK2). We may notice that MP5 provides a more accurate capturing of the various waves and discontinuities, in spite of the same overall maximum order achieveable, with some extra oscillations, which are anyway damped at higher resolutions, as in Fig.~\ref{fig:st}. Spurious oscillations (Gibbs phenomena) near shocks are a well known price to pay for high order schemes, especially for those avoiding decomposition in characteristics, like ECHO. However, we deem that the post-processing MP filter behaves quite well in this kind of tests.

\subsection{Cylindrical blast wave}

Let us treat RMHD problems involving shocks in more than one dimension. A notoriously hard test for relativistic codes is the cylindrical blast wave expanding in a plasma with an initially uniform magnetic field. This problem was already considered in Paper~II, here we test our new MP5 scheme and we adopt the more widely used settings by \citet{komissarov99}. Unfortunately, no exact solution is available for the present problem. From a numerical point of view, in the multidimensional relativistic case it is very difficult to treat correctly situations with flow of Alfv\'en velocities close to the speed of light. This is because the numerical errors, which are always present in the reconstruction procedures, act \emph{independently} on, say, $x$ and $y$ components of $\vec{v}$ and $\vec{B}$ in 2-D runs. This problem easily leads to uncorrect fluxes and eventually provides unphysical states, e.g. with $v^2>1$, when primitive variables are recovered from the evolved conservative ones. Moreover, terms in the total energy equation are strongly unbalanced in these cases and, again, numerical errors may lead to code crashing.

The initial conditions are as follows: a square Cartesian box $[-6,6]\times [-6,6]$ contains an internal cylindrical region, within $r=(x^2+y^2)^{1/2}\le 1$, with $\rho=10^{-2}$ and $p=1$. This region is surrounded by an external medium with $\rho=10^{-4}$, $p=5\times 10^{-4}$ and these values are reached by means of a smooth ramp function betwen $r=0.8$ and $r=1$. The velocity is zero everywhere and the magnetic field is uniform, with $B_x=0.1$. This is the intermediate magnetization case by Komissarov, with a higher external pressure as in \citet{leismann05}. We are not able to run this test with stronger fields or lower external pressure without introducing \emph{ad hoc} numerical strategies. In Fig.~(\ref{fig:blast2d}) we show several quantities at $t=4.0$, for a run with $200\times 200$ grid points. The scheme used is, as in the previous test, MP5 for REC, no DER, and RK2 for time-stepping (overall second order in both space and time). We notice the presence of several structures: an external fast shock, an inner region bounded by a reverse shock, both almost circular, and complex anisotropic discontinuities in between. Note, in particular, that the magnetic field is almost completely swept out from the central region by the explosion. The highest outflow speed is reached for $y=0$ ($\Gamma_\mathrm{max}=3.69$), since there is no magnetic force preventing the expansion in the direction along the fieldlines. This problem is also a severe test because of the various degeneracies which may occur in the Riemann solver. In our case, the HLL procedure with the simplified calculation of fast wave speeds does not suffer this kind of problems. In spite of the simplified Riemann solver, structures appear well defined thanks to the use of an accurate REC routine.

\begin{table}
\centering
\begin{tabular}{lcccc}
\hline\hline
Method & CPU time & iter. & iter. / time & sub-iter. / time \\
\hline
MC2-RK2   &  86.6 $s$ & 133 & 1.54 $s^{-1}$ & 3.08 $s^{-1}$ \\
WENO5-RK2 &  97.6 $s$ & 132 & 1.35 $s^{-1}$ & 2.70 $s^{-1}$ \\
MP5-RK2   & 100.1 $s$ & 132 & 1.32 $s^{-1}$ & 2.64 $s^{-1}$ \\
WENO5-RK3 & 152.9 $s$ & 132 & 0.86 $s^{-1}$ & 2.58 $s^{-1}$ \\
MP5-RK3   & 154.4 $s$ & 133 & 0.86 $s^{-1}$ & 2.58 $s^{-1}$ \\
\hline
\end{tabular}
\caption{Efficiency results for the cylindrical blast wave problem. CPU time (in seconds), total number of iterations, iterations per second, and sub-RK cycles per second are reported for various schemes (the DER routine is used only in schemes adopting RK3).}
\label{tab:blast2d}
\end{table}

As far as efficiency is concerned, we use the present test to measure the CPU time for different scheme settings of ECHO. Results are reported in Table~(\ref{tab:blast2d}), where data refer to double precision runs on an Intel Xeon 3.0 Ghz processor, for Linux operating system, with the Intel Fortran compiler. The best performing scheme is obviously that based on linear reconstruction, MC2 in this case, whereas MP5-RK3 is 1.78 times slower. However, as we can see from the sub-cycles per second, it is the order of the Runge-Kutta method that matters most, whereas the DER procedure is quite efficient. When comparing reconstruction schemes of the same order, we can notice that MP5 is just slightly slower than WENO5 in our implementation, probably due to the minmod-type conditions in the limiting process. However, from our tests we have found that MP5 is both more accurate for smooth solutions and more robust (less oscillatory) in problems involving shocks. Our conclusion is that MP5 employed at an overall spatial and temporal second order gives the best trade-off among efficiency, accuracy and robustness, thus it will be used as our base scheme in the next numerical tests.

\subsection{Radial accretion in Schwarzschild metric}
\label{sect:michel}

\begin{figure}
\centering
\resizebox{\hsize}{!}{\includegraphics{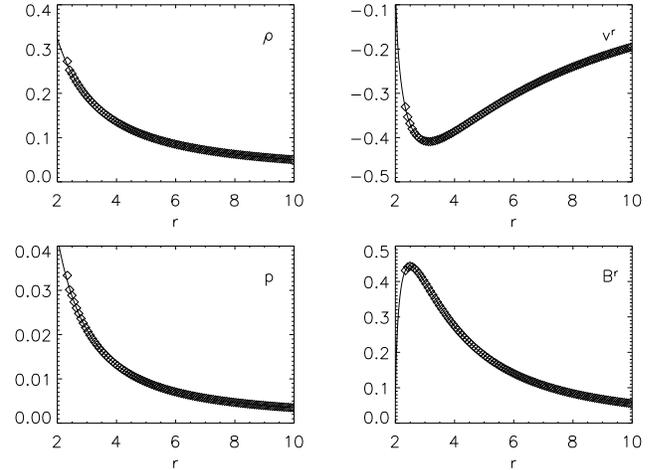}}
\caption{
Results for the 1-D accretion flows in Schwarzschild metric. Quantities are shown by plotting the numerical results at $t=100$ (diamonds) over the respective exact solution (solid line). A resolution corresponding to 100 grid points and reconstruction with MP5 are used.
}
\label{fig:michel}
\end{figure}

\begin{figure}
\centering
\resizebox{\hsize}{!}{\includegraphics{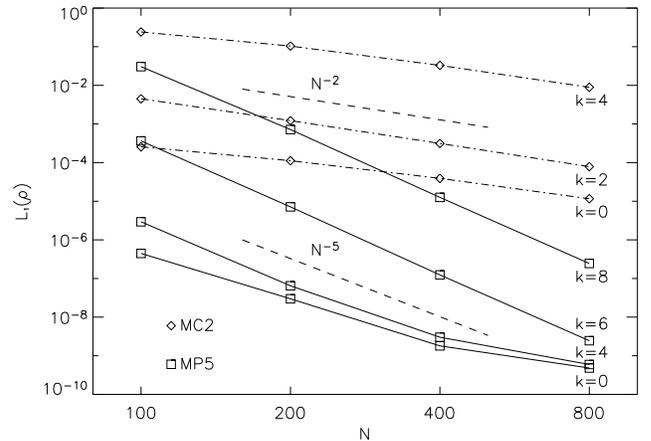}}
\caption{
Errors for the 1-D accretion flow in Schwarzschild metric of Fig.~(\ref{fig:michel}). The $L_1$ norm of the density is shown as a function of the grid points $N$, for MC2-RK2 (diamonds) and MP5-RK3 (squares). The plasma beta at $r_c$ is also varied, according to the parameter $k=-\log_{10}{\beta_c}$.
}
\label{fig:michel_error}
\end{figure}

As a first test in a curved space-time we consider here the spherical transonic accretion onto a non-rotating black hole (of mass $M=1$) in the presence of a radial magnetic field. The aim is to check the code ability to preserve in time an analytical solution in a curved geometry, where metric terms and their derivatives are involved. A full description of the (fluid) transonic stationary solution is given in \citet{michel72}, here we follow the setup of \citet{anton06}. We hence adopt Schwarzschild metric and coordinates, with a singular horizon for $r_h=2$, where the lapse function $\alpha=(1-2/r)^{1/2}$ vanishes and $\gamma_{rr}=\alpha^{-2}$ diverges. The numerical domain is $2.3<r<10$, the critical point radius is $r_c=8$, an isentropic condition is assumed, and the remaining free constants are chosen by setting $\rho_c=1/16$ (in order to have a mass flux of $r^2\rho\Gamma v^r=-1$) and by assigning the value of the plasma beta at the critical radius, $\beta_c=2p_c/B_c^2$, which we leave as a free parameter. Note that from an analytical point of view the Michel solution does not change in the presence of a monopole magnetic field, thus the fluid quantities are unaffected by the value of the plasma beta (only the magnetic field $B^r$ will depend on it, namely as $\beta_c^{-1/2}$), whereas numerically the presence of a large magnetic field may lead to severe errors and code breaking. This is mainly due to the fact that the numerical derivatives of magnetic terms in fluxes do not balance exactly the corresponding source terms in the momentum equation, and secondly because of the difficulties encountered in the inversion routine for the primitive variables.

In Fig.~(\ref{fig:michel}) we show the results of a simulation with $\beta_c=1$ and $N=100$ grid points in the radial direction, comparing the quantities obtained at $t=100$ with the analytical solutions. The scheme employed is MP5 at second order of overall accuracy. Small discrepancies can be seen only near the inner radius, where gradients are the largest. To remove both these large gradients and the singularity at $r=2$ horizon-adapted coordinate systems could also be used \citep{papadopoulos98}, but here we prefer to use the standard Schwarzschild coordinates. For a more quantitative comparison, we report in Fig.~(\ref{fig:michel_error}) the normalized $L_1$ errors of the density as a function of the grid points from $N=100$ to $N=800$, for the two schemes MC2-RK2 and MP5-RK3 (here with DER). The value of the plasma beta is also varied, from $1$ to $10^{-8}$ (for an increasing magnetization $\sigma=B^2/\rho$, approximately from $10^{-1}$ to $10^7$). The first thing to notice is that the expected scaling with $N$ works also in this non-Cartesian case (though there is the usual saturation effect around $10^{-10}$). Then we see that the Runge-Kutta order is not an issue in this kind of test, where stationary flows are involved (otherwise the maximum order would have been 3). Moreover, high resolution schemes allow us to reach much lower betas (for $N=800$ down to $\beta_c=10^{-8}$ with MP5 at full spatial accuracy order $r=5$, and $\beta_c=10^{-6}$ with MC2). If MP5 is employed at second order, intermediate results are found (not reported in the plot). In order to be able to reach such low plasma betas, we have here used Eq.~(\ref{eq:adiabatic}), the adiabatic equation for the entropy function $s=p/\rho^\gamma$ (the solution is smooth). If the full energy equation is used errors are larger by a factor $\approx 2$.

\subsection{Equatorial accretion in Kerr metric}
\label{sect:gammie}

As another example of 1-D test in a curved space-time, we proceed further in the level of complexity by studying an accretion problem in Kerr metric, where not only the lapse function $\alpha$ is involved, but also the shift vector $\vec{\beta}$. The problem is the magnetized equatorial flow in Kerr metric described by \citet{takahashi90}. It is basically the general relativistic analog of the \citet{weber67} model for the solar wind, where the radial velocity has to pass smoothly three critical points (slow, fast and Alfv\'enic) in the equatorial plane where the Parker spiral of magnetic field lies. The accretion solution was later specialized to the region between the black hole horizon and the marginally stable orbit by \citet{gammie99}, in which a cold inflow has to cross just the Alfv\'enic critical point (coincident with the magnetosonic fast point for vanishing thermal pressure). For our numerical test we use the settings proposed by \citet{gammie03} and \citet{devilliers03a}, that is we study the accretion onto a Kerr black hole with $a=0.5$, which gives an event horizon at $r_h=1+(1-a^2)^{1/2}\simeq 1.866$ (the spherical surface where $\gamma_{rr}$ diverges) and a marginally stable orbit at $r_\mathrm{mso}\simeq 4.233$. After choosing the other free parameters, the critical point is located at $r_c\simeq 3.617$. The pressure is initialized with an isentropic law, preserving a vanishing thermal contribution $p\ll\rho\Rightarrow h\simeq 1$.

\begin{figure}
\centering
\resizebox{\hsize}{!}{\includegraphics{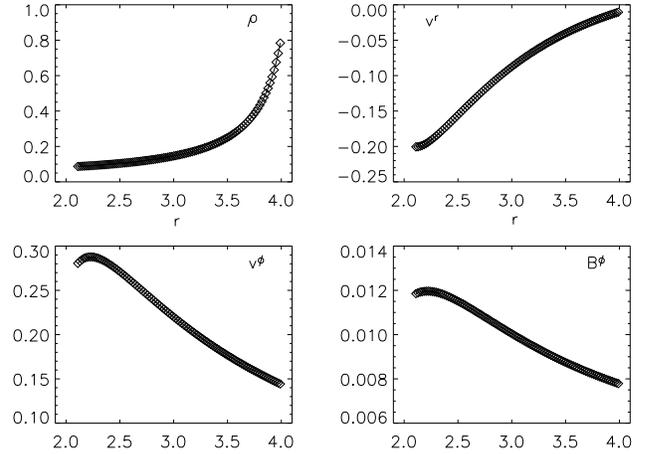}}
\caption{
Results for the 1-D accretion flow in Kerr metric. Quantities are shown by plotting the numerical results at $t=100$ (diamonds) over the respective exact solution (solid line). A resolution corresponding to 100 grid points and reconstruction with MP5 are used.
}
\label{fig:gammie}
\end{figure}

In this test we adopt Boyer-Lindquist coordinates and a radial domain $2.1<r<4.0$ with $N=100$ grid points. The results are shown in Fig.~(\ref{fig:gammie}), where the significant physical quantities are plotted at the output time $t=100$ against the initial solution. As in the previous case, at the outer boundary, where the inflow is originated, all quantities are kept constant in time. The scheme employed is MP5 at overall second order (no DER and RK2), which is rather accurate already at this low resolution, even at the inner boundary which is close to the event horizon.  For a quantitative comparison with the other reconstruction schemes, we report here the $L1$ errors on the normalized density as in the previous section, again for $N=100$ and $t=100$. MC2 gives 2.76e-3, CENO3 ($r=3$ and RK3) gives 2.42e-4, both MP5 and WENO5 ($r=5$ and RK3) give 1.40e-4, while MP5 at second order gives 3.22e-4. The improvement of high order methods is not as apparent as in the previous test, due to limited precision in the initializing routines.

\subsection{Axisymmetric torus in Kerr metric}

\begin{figure*}
\centering
\resizebox{\hsize}{!}{
\includegraphics{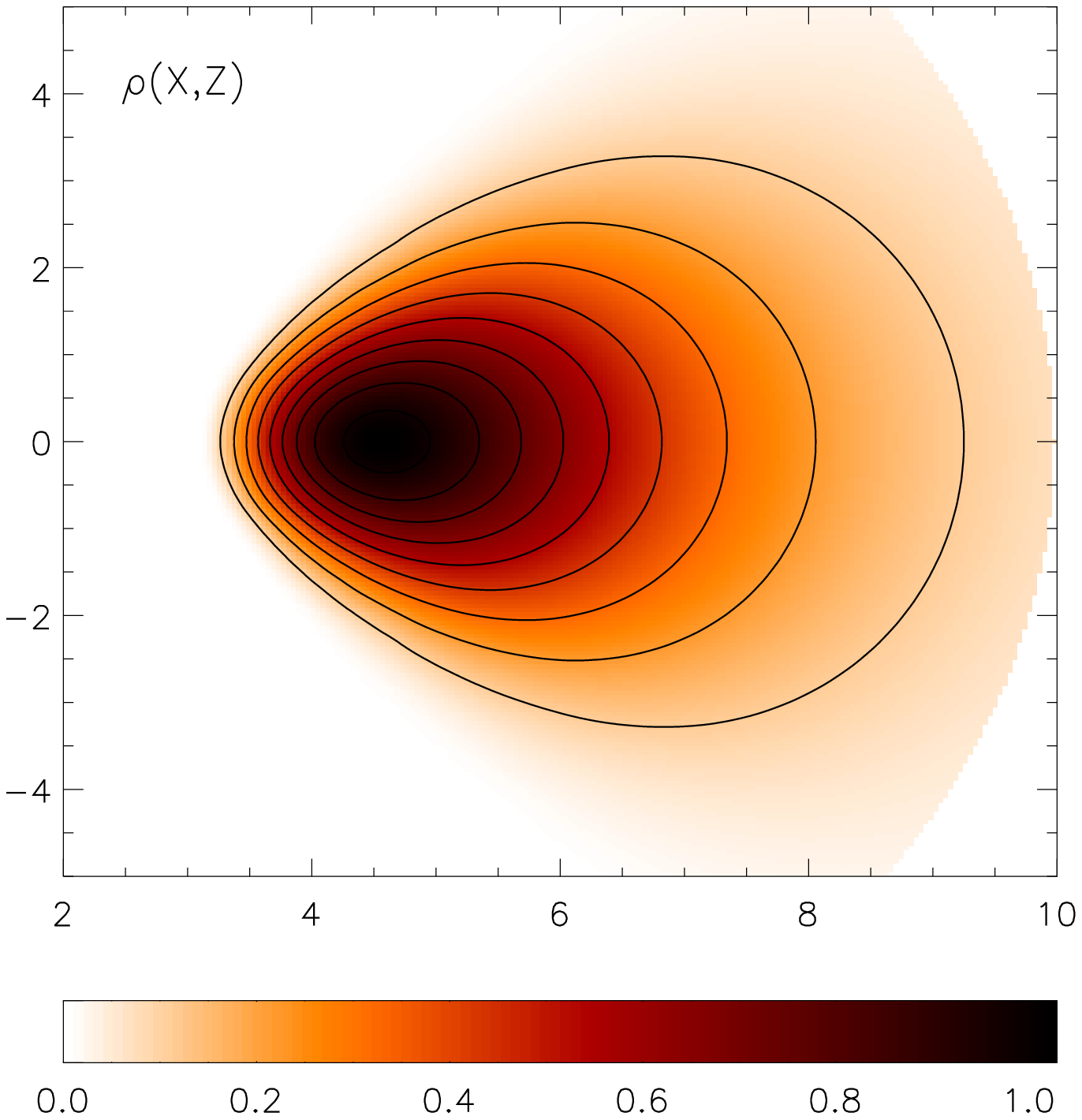}
\includegraphics{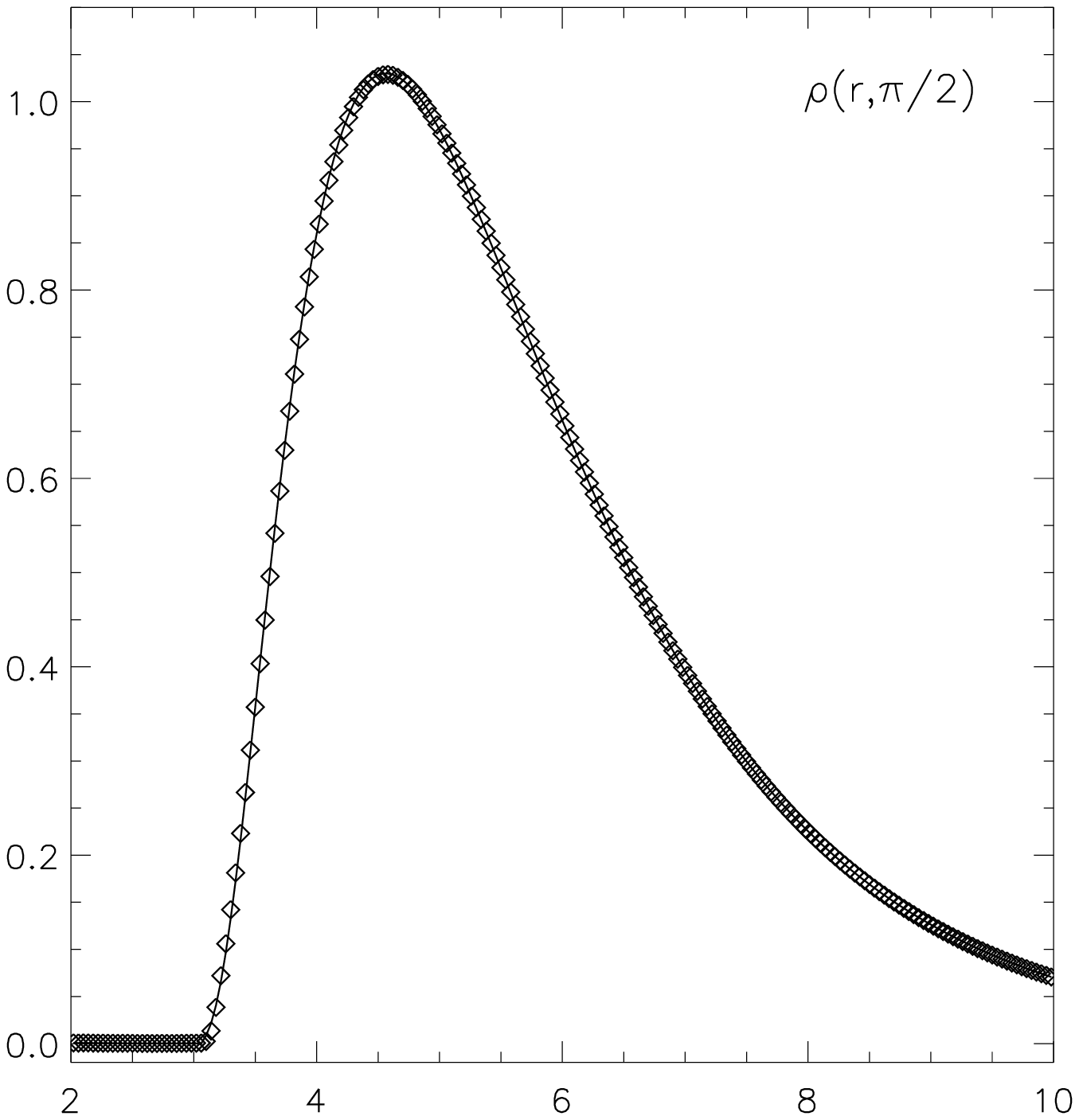}
\includegraphics{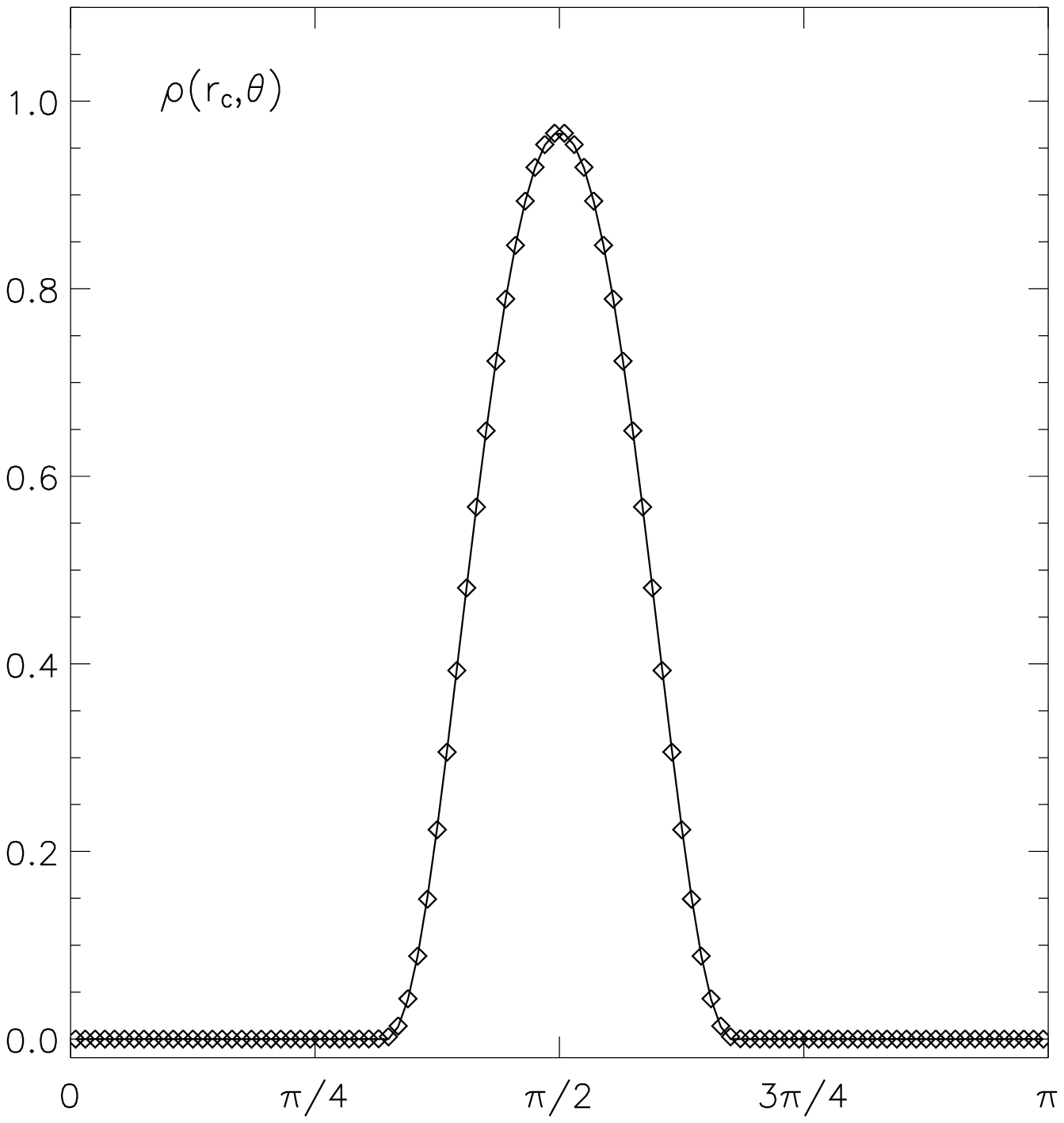}
}
\resizebox{\hsize}{!}{
\includegraphics{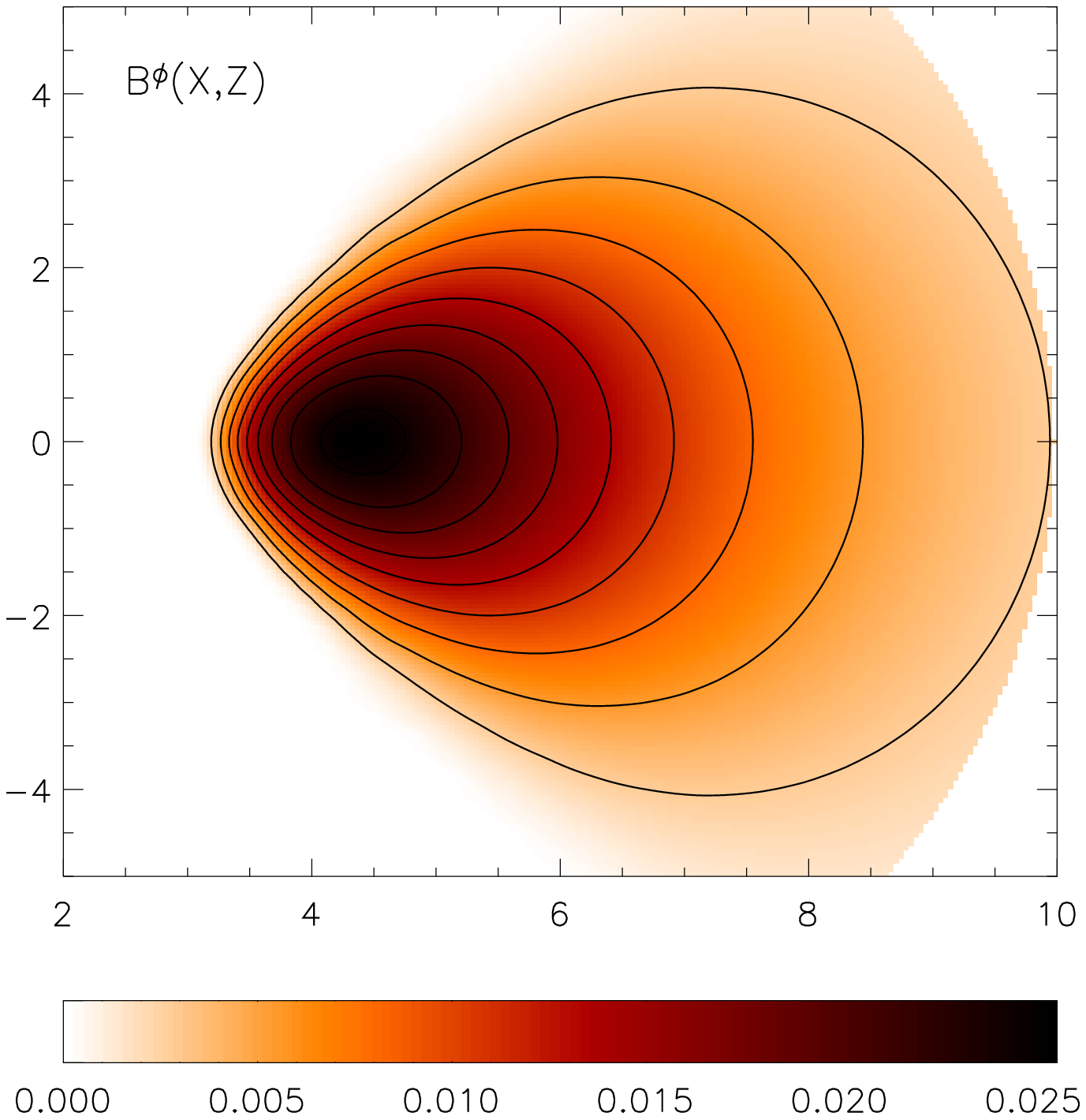}
\includegraphics{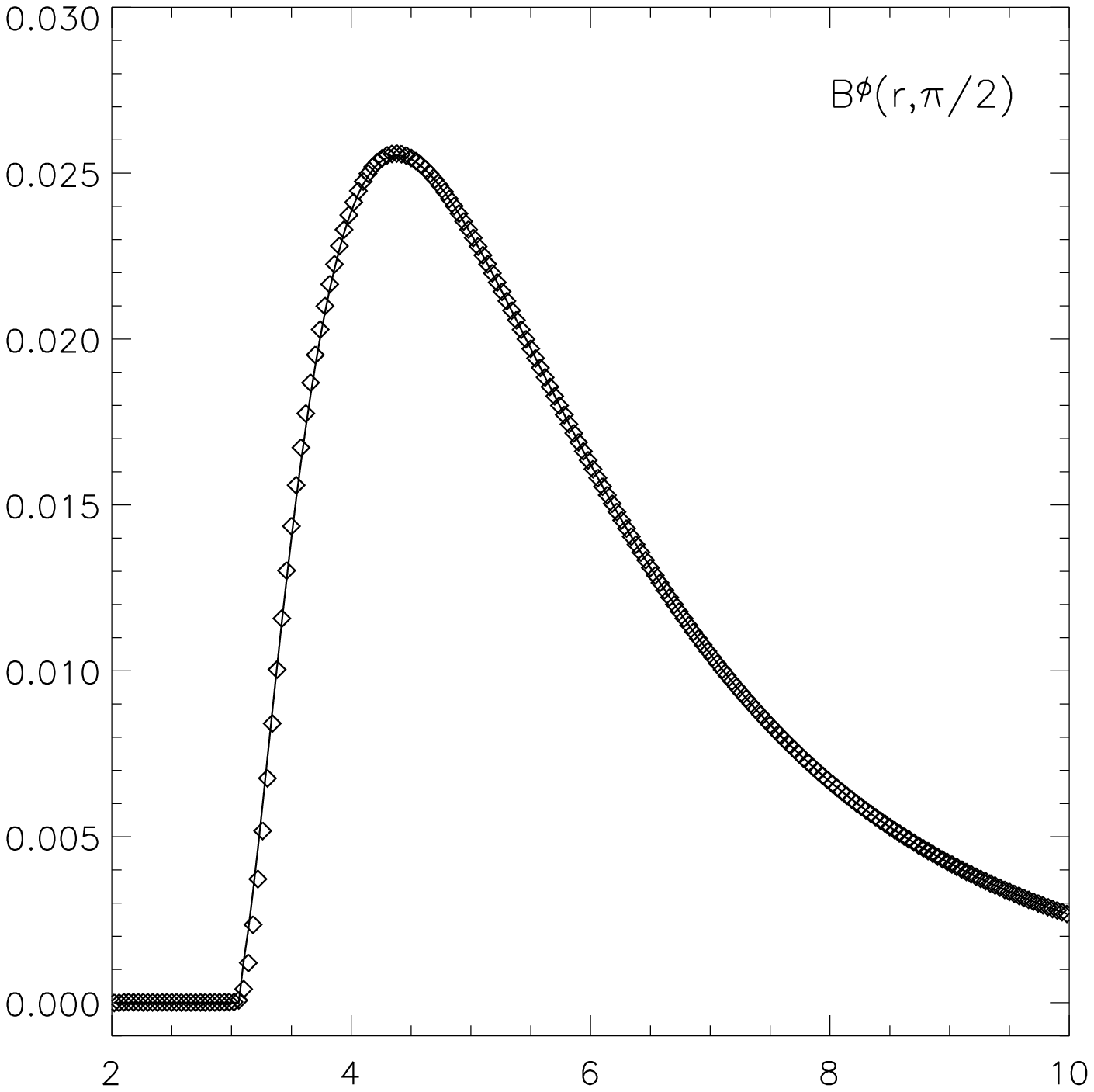}
\includegraphics{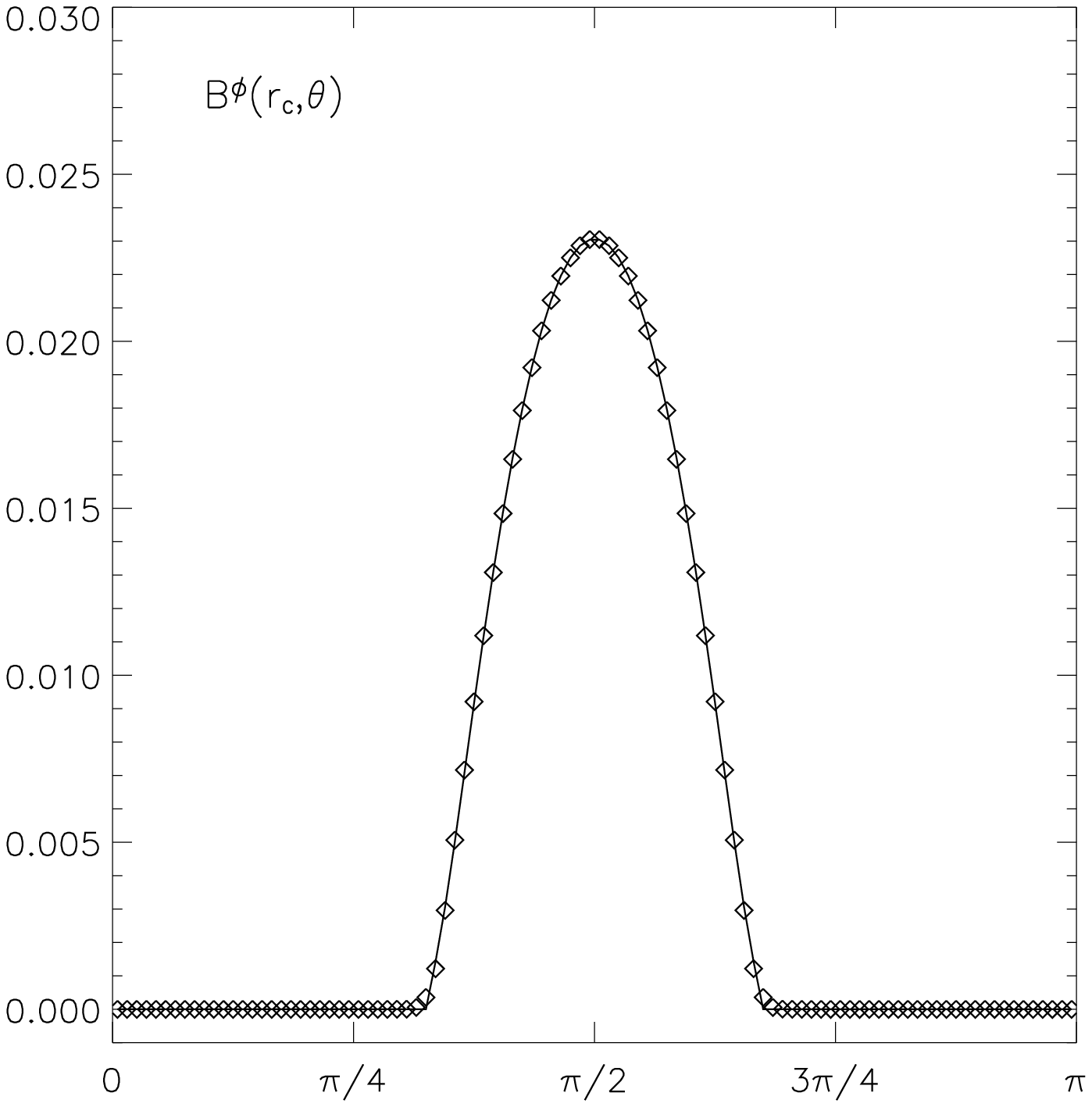}}
\caption{
The results of the magnetized disk evolution. Density (upper panels) and toroidal field (lower panels) are displayed at the final output time $t=200$. 2-D maps (cylindrical coordinates $X=r\sin\theta, Z=r\cos\theta$ are used for ease of graphical presentation), radial and latitudinal cuts through the disk center $(r_c,\pi/2)$ are shown for the two quantities. Solid lines in the 1-D profiles refer to the initial analytical solutions.
}
\label{fig:disk}
\end{figure*}

The final GRMHD test proposed here is to study the stability of a constant angular momentum thick disk around a Kerr black hole and threaded by a toroidal magnetic field. This will be achieved through simulations in a 2-D domain, assuming axisymmetry. For the analytical theory of equilibrium in the purely hydrodynamical case the reader is referred to \citet{abramowicz78,kozlowski78,font02}, while the GRMHD version with the addition of a purely toroidal magnetic field is due to \citet{komissarov06c}. This test may also represent the basis for studying a class of relevant astrophysical problems, since the dynamics of accretion disks orbiting around black holes is believed to be strongly influenced by the presence of magnetic fields. We summarize here the main features of the equilibrium model, while addressing to \citet{komissarov06c} for a more detailed description. 
Under the assumptions of purely toroidal velocity and magnetic field, the Bernoulli-like equation that needs to be solved is 
\be
\label{eq:disk}
\mathrm{d}\ln (-u_t)-\frac{\Omega\,\mathrm{d}\ell}{1-\ell\Omega}+\frac{\mathrm{d}p}{\rho h}+\frac{\mathrm{d}(R^2p_\mathrm{m})}{R^2 \rho h}=0,
\ee
where $\ell=-u_\phi/u_t$ is the specific angular momentum, $\Omega=u^\phi/u^t$ is the angular velocity, $p_\mathrm{m}=B^2/2$ is the magnetic pressure (notice that the electric field vanishes since $\vec{v}\parallel\vec{B}$), and $R^2=(g_{t\phi})^2-g_{tt}g_{\phi\phi}$ is the generalized distance from the rotation axis. We then assume a constant distribution of the specific angular momentum, i.e. $\ell=\ell_0$, such that Eq.~(\ref{eq:disk}) provides the potential
\be
W=\ln(-u_t)=\frac{1}{2}\ln\left(\frac{R^2}{g_{\phi\phi}+2g_{t\phi}\ell_0+g_{tt}\ell_0^2}\right).
\ee
The equation of state is barotropic and it is convenient to choose $p\propto (\rho h)^\gamma$ for the thermal contribution and similarly $R^2p_\mathrm{m}\propto (R^2 \rho h)^\gamma$ for the magnetic pressure. Under these assumptions Eq.~(\ref{eq:disk}) can be integrated as
\be
W-W_\mathrm{in}+\frac{\gamma}{\gamma-1}\frac{p+p_\mathrm{m}}{\rho h}=0.
\ee
The disk is characterized by the condition $W\le W_\mathrm{in}$, where $W_\mathrm{in}$ is calculated at the inner disk radius $r_\mathrm{in}$ on the equatorial plane. The cusp and the center of the disk are defined as those points, again in the equatorial plane, where the specific angular momentum retains its Keplerian value. Here we use the radius of the disk center, $r=r_c$, to determine $\ell_0$. Notice that the potential $W$ has a local minimum at $r_c$, though only in the purely hydrodynamical case this point also corresponds to the maxima of $\rho$ and $p$. The overall disk structure is then completely specified by the two radii $r_\mathrm{in}$ and $r_c$, and by the density $\rho_c$ and plasma beta $\beta_c$ at the disk center.

Outside the disk ($W>W_\mathrm{in}$) we define a static, unmagnetized atmosphere in equilibrium with gravity. This can be obtained by adopting the solution of the relativistic Bernoulli equation for an isentropic plasma $p\propto\rho^\gamma\Rightarrow\mathrm{d}p/(\rho h)=\mathrm{d}\ln h$, and Eq.~(\ref{eq:disk}) is readily integrated to give the simple relation
\be
h\, (-u_t)=h\,(R^2/g_{\phi\phi})^{1/2}=\mathrm{const}.
\ee
The exact solution can be determined by providing the values $\rho_\mathrm{atm}$ and $p_\mathrm{atm}$, calculated for example at the disk center. Notice that all the above relations are valid for both Boyer-Lindquist coordinates and Kerr-Schild coordinates \citep[e.g.][]{komissarov04a}, which have non-vanishing $g_{tr}$, $g_{t\phi}$ and $g_{r\phi}$ terms needed to remove the (unphysical) singularity at the event horizon. In the case of Boyer-Lindquist coordinates, employed here for the numerical test, we have $R^2=\alpha^2g_{\phi\phi}\Rightarrow -u_t=\alpha$, so that the equilibrium condition for the static atmosphere is simply $\alpha h=\mathrm{const}$. A more physical option for the external environment would be to define the spherically symmetric Michel's transonic inflow and let the system relax to steady state. However, since here we are mainly interested in the stability of the disk itself, we prefer to use the above static solution, which is an exact one and it is much simpler to be initialized.

The simulation setup is as follows. The numerical domain is taken to be $2<r<10$ and $0<\theta<\pi$, with 200 grid points in the radial direction and 100 in the polar angle direction. We keep the quantities fixed in time at both radial boundaries, while reflecting conditions are imposed at the poles. The first condition is needed because otherwise numerical errors near the inner radial boundary, where the gradients are the largest, tend to destabilize the whole atmosphere. This problem could be also cured by choosing appropriate non-uniform grids with higher resolution at small radii and/or by using Kerr-Schild coordinates, but here we want to retain the simplest possible test conditions. The free parameters are chosen to be $a=0.99$, $r_\mathrm{in}=3$, $r_c=5$, $\rho_c=1$, $\beta_c=1$, $\rho_\mathrm{atm}=10^{-5}\rho_c$, $p_\mathrm{atm}=0.1\rho_\mathrm{atm}$. Note that the value of $\rho_c$ is arbitrary since we are not evolving the metric, which is determined by the central black hole mass (here taken as unity) and angular momentum alone. With the present values we find $\ell_0\simeq 2.80$, $W_\mathrm{in}\simeq -4.16\times 10^{-2}$, $W_c\simeq -9.83\times 10^{-2}$ so that the inner disk disk is located beyond the cusp point and there is a finite outer radius (beyond the computational domain). The rotation period at the disk center is $2\pi/\Omega_K(r_c)=2\pi (r_c^{3/2}+a)\simeq 76.5$, we take $t=200$ as the final output time, corresponding to just a few orbital periods but a much longer time with respect to the local dynamical timescales. Here we use MP5 reconstruction at an overall $r=2$ accuracy (no DER and RK2 for time integration).

The results are shown in Fig.~(\ref{fig:disk}), where in the upper panels we show the density (2-D map, radial cut through the disk center, latitudinal cut through the disk center) and in the lower panels the toroidal field $B^{\phi}$. In the maps we show color images and contours of the quantities evolved at the final time (indistinguishable from those at $t=0$), whereas for the 1-D cuts we plot the numerical solution at $t=200$ (diamonds) together with the initial conditions (solid line). Note that our reconstruction scheme based on MP5 behaves very well. Minor discrepancies appear only near the steep boundaries between the disk and the external atmosphere (where density jumps of order $10^3-10^4$ are initially captured by just 2-3 points). Angular momentum is also transferred to the non-rotating external atmosphere due to numerical diffusion in the vicinities of the disk boundaries. The $L_1$ norm is 2.8e-4 for the density and 2.41e-5 for the magnetic field, while the $L_\infty$ norm (the largest error in absolute value) is 1.60e-2 and 1.31e-3, respectively. These results appear to be comparable to those presented by \citet{komissarov06b}, in spite of the use of a much simpler Riemann solver, a constant radial grid spacing, and retaining the same overall second order accuracy. Finally, errors around the disk boundaries due to numerical diffusion are much larger if MC2 is employed instead of MP5, confirming that reconstruction based on large stencils may help even near discontinuities. The situation is improved if a non-linear radial grid is employed, in that case also MC2 provides a good accuracy. On the other hand, results obtained with the full fifth order scheme (and RK3 for time-stepping) are similar for this case. Finally, note that the present test has been performed by solving the full energy equation, and no appreciable changes are noticed when Eq.~(\ref{eq:adiabatic}) is solved instead. 

\section{GRMD numerical tests}
\label{sect:grmd_tests}

In the present section we perform a series of tests to check the performances of ECHO when configured for special and general relativistic magnetodynamics. The numerical settings are the same as in the base scheme used for the GRMHD tests, namely we employ the HLL solver coupled to MP5 for the reconstruction (switching off the additional corrections to achieve effective higher accuracy) and RK2 for time integration.

\subsection{Propagation of waves and discontinuities}
\label{sect:waves}

\begin{figure}
\centering
\resizebox{\hsize}{!}{\includegraphics{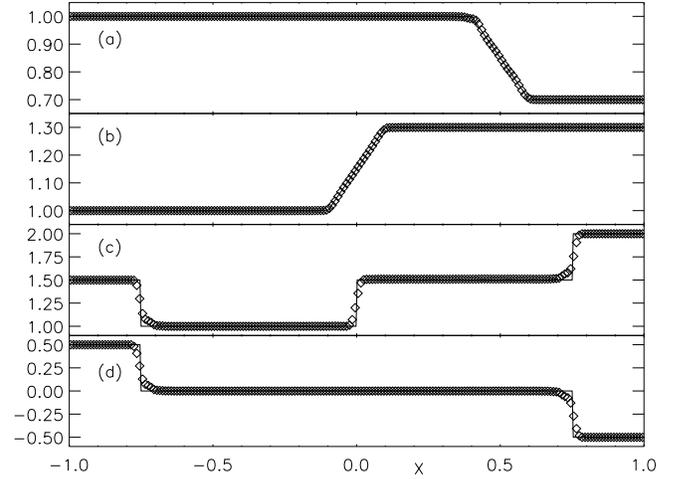}}
\caption{
The set of four magnetodynamics 1-D test problems selected in Sect.~\ref{sect:waves}. All plots refer to the $B^y(x)$ transverse field component at the final output time corresponding to each test. From above to below, the four test problems are: fast wave, stationary Alfv\'en wave, three-waves, current sheet.
}
\label{fig:waves}
\end{figure}

Several 1-D tests have been proposed for special relativistic MD. Here we select four of them and we change slightly some of the original setups found in the literature in order to make the notation more uniform. In all runs we assume a numerical domain of $200$ grid points in the interval $-1.0\leq x\leq 1.0$ and a constant background field $B^x=1.0$. The results of the corresponding simulations, at different output times, are all plotted in Fig.~\ref{fig:waves}, where the transverse component $B^y(x)$ is shown.

\begin{itemize}

\item (a) \emph{Fast wave}. Here $B^z=E^x=E^y=0.0$ and, following \citet{komissarov02}, the transverse magnetic field component is
\be
B^y(x)=\left\{ \begin{array}{ll} 
1.0, & x<-0.6 \\
1.0-1.5 (x+0.6), & -0.6<x<-0.4 \\
0.7, & x>-0.4,
\end{array}\right.
\ee
whereas $E^z(x)=1-B^y(x)$. The fast wave is initially centered at $x=-0.5$ and then should propagate with unchanged profiles at the speed of light. We use $t=1.0$ as output time, so that the final position will be $x=0.5$. Some small wiggles are barely visible in the numerical solution near the corners of the wave profile, otherwise the agreement with the analytical solution is very good.

\item (b) \emph{Stationary Alfv\'en wave}. An initial setting similar to that in \citet{komissarov04a} is assumed for this test, though we swap the role of the transverse electromagnetic components and the wave profile to make it more similar to the previous test. Here we take $B^z=E^y=1.0$, $E^z=0.0$, and
\be
B^y(x)=\left\{ \begin{array}{ll} 
1.0, & x<-0.1 \\
1.0+1.5 (x+0.1), & -0.1< x<0.1 \\
1.3,  & x>0.1,
\end{array}\right.
\ee
now with $E^x(x)=-B^y(x)$. This solution is a stationary linearly polarized MD Alfv\'en wave centered at $x=0$, so its profile should be preserved in time and only numerical dissipation effects should be found. The output time used for this test is $t=2$ and from the plot we can see that numerical dissipation is negligible for this test, as the initial and final profiles are indistinguishable.

\item (c) \emph{Three-waves}. This test was proposed by \citet{komissarov02} and it is concerned with the splitting of a discontinuity initially located at $x=0$ into three waves: two oppositely propagating fast waves (traveling at the speed of light) and a standing Alfv\'en wave. It is thus a MD analogue of a RMHD shock tube, with the only difference that shocks are not allowed in the MD limit. The initial conditions are
\be
(\vec{B},\vec{E})=\!\left\{ \begin{array}{llllrrl} 
(1.0, & 1.5, & 3.5, & \!-1.0, & \!-0.5, & \! 0.5), & x<0, \\
(1.0, & 2.0, & 2.3, & \!-1.5, & \! 1.3, & \!-0.5), & x>0,
\end{array}\right.
\ee
and the output time is $t=0.75$. From the plot we can see that the fast wave fronts are reasonably sharp and the the Alfv\'enic discontinuity is preserved within only four grid points. The combination of our simple two-waves HLL solver with high resolution reconstruction methods like MP5, even when employed in an overall second order scheme, confirms thus its validity in this kind of tests. Note in particular the absence of spurious oscillations, a possible drawback of reconstruction methods based on large stencils in problems with sharp discontinuities, which proves the limiting capabilities of the monotonicity preserving algorithm.

\item (d) \emph{Current sheet}. A current sheet is easily set up by choosing $B^z=E^x=E^y=E^z=0.0$ and
\be
B^y(x)=\left\{ \begin{array}{rl} 
 0.5, & x< 0, \\
-0.5, & x> 0,
\end{array}\right.
\ee
as in \citet{komissarov04a}. With this value of the transverse field, the constraint $B^2-E^2\geq 0$ is easily preserved throughout the evolution. At the output time $t=0.75$ we can see in the figure the two oppositely propagating fast wave fronts located at $x=\pm 0.75$, as expected. The numerical diffusion of these shocks is very similar to that in the previous test.

\end{itemize}

\subsection{Uniform magnetic field in Schwarzschild metric}

\begin{figure}
\centering
\resizebox{\hsize}{!}{\includegraphics{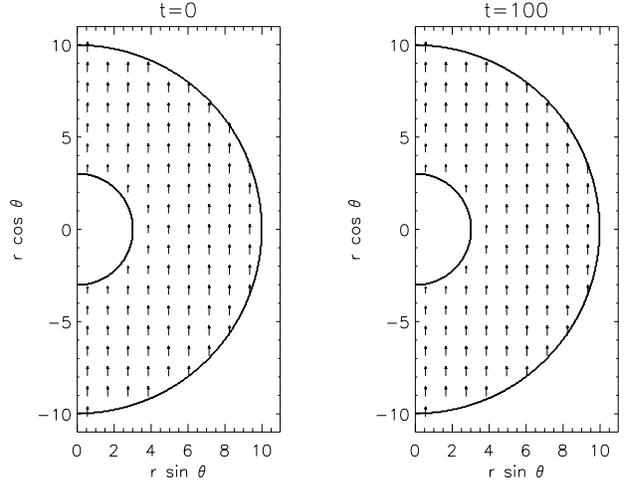}}
\caption{
The Wald solution in Schwarzschild metric at the initial time $t=0$ and output time $t=100$. The code employs Boyer-Lindquist coordinates, here cylindrical coordinates $X=r\sin\theta, Z=r\cos\theta$ are used for ease of graphical presentation.
}
\label{fig:wald}
\end{figure}

As a 2-D test in a curved space-time we consider the equilibrium force-free solution found by \citet{wald74}, here in Schwarzschild metric. An exact solution for the magnetic field is
\be
B^r=B_0\,\alpha\cos\theta,~~~B^{\theta}=-B_0\,\alpha r^{-1}\sin\theta,
\ee
whereas $B^{\phi}=0$ and $\vec{E}=0$. When translated into cylindrical coordinates ($R=r\sin\theta$, $Z=r\cos\theta$), this is a uniform vertical field of strength $B_0$ aligned with the $Z$ axis. For our test we choose $B_0=1$ and a numerical domain $3\leq r\leq 10$, $0\leq\theta\leq\pi$, with $200$ grid points in the radial direction and $100$ in the latitudinal direction. The initial equilibrium is evolved to a large time $t=100$ (the light crossing time in the radial direction is $t=7$) and in Fig.~\ref{fig:wald} we report the magnetic field in vectorial form (the length of the arrow is proportional to its strength) for $t=0$ and $t=100$. Only minor discrepancies are visible, for an average error of $\approx 6\times 10^{-3}$ in the field strength.

\section{Conclusions}
\label{sect:concl}

We have presented a new code, ECHO, that is the extension of our central-type special relativistic scheme (Paper~I and II) to general relativistic MHD and magnetodynamics. This is achieved by applying the general UCT strategies \citep{londrillo00,londrillo04} for MHD-like hyperbolic systems of conservation laws. The resulting numerical scheme is based on simplified Riemann solvers and finite difference high order reconstruction methods. As far as the general relativistic framework is concerned, we adopt here the so-called $3+1$, or Eulerian, formalism. This allows us to present the equations (in conservative form) in the most familiar way, i.e. resorting to three-dimensional vectors and tensors alone. The limits to special relativistic MHD and classical MHD are then straightforward in this framework. Gravitational terms appear in fluxes and in the external sources avoiding the use of complex four-dimensional Christoffel symbols. The metric can also depend on time and be provided by any solver for Einstein's equations.

ECHO's high order procedures are first tested in flat space-time, with a new problem involving the propagation of \emph{large amplitude} Alfv\'en waves in 1-D and 2-D domains. We demonstrate that the same settings valid for classical MHD can be employed in the RMHD problem too by only changing the propagation speed. This now depends on the amplitude of the wave itself, due to the electromagnetic energy contribution to the overall inertia. For the reconstruction routines tested the nominal high order of overall accuracy is always reached, up to fifth order. For the same problem, spectral properties are also checked for various schemes by investigating the code behavior at small wavelengths, where second order schemes usually fail. Moving to discontinuous solutions, one magnetized shock tube is tested and even in this case our reconstructions based on larger stencils seem to provide sharp profiles also on contact-type discontinuities, where approximate Riemann solvers usually give poor results. In 2-D we study the magnetized blast wave problem, where difficulties are known to arise when Cartesian grids are used. We find that when the Lorentz factor and/or the magnetization are too high, then numerical errors (which are independent along each direction) may lead the code to crash.

In curved space-times, we first study the radial accretion onto Schwarzschild black holes in the presence of a monopole magnetic field. High order schemes are able to reproduce the analytical solution much better, and this allows us to reach a magnetization as high as $10^7$ (for typical values of the other parameters), while TVD-like second order schemes usually start to fail around $10^2-10^3$. The expected scaling with the accuracy order is also reproduced for this test in non-Minkowskian metric. In Kerr space-time we test the 1-D equatorial accretion and the 2-D stability of a constant angular momentum thick disk with a toroidal magnetic field (a recently obtained exact solution). The latter test provides a very important astrophysical scenario, since magnetized tori and rotating black holes are the likely ingredients for AGN and microquasar energy release. Our scheme with limited reconstruction based on a five-point stencil is able to maintain the equilibrium solution for several rotation periods with negligible errors. This result is achieved without adopting specifically designed non-linear grids or horizon adapted coordinate systems, thus proving its robustness in complex situations of astrophysical interest.

With only minor modifications the GRMHD scheme has been also tested in the force-free, low-inertia limit of (ideal) magnetodynamics \citep{komissarov02}. The fluid velocity is replaced by the drift velocity of magnetic field lines and the same conservative approach is kept unaltered \citep{mckinney06a}. We then study the propagation of MD waves and discontinuities in flat space-time and the stability of a uniform magnetic field around a Schwarzschild black hole.

As far as efficiency is concerned, a scheme which is fifth order accurate in space and third order in time is $\approx 1.8$ times slower than a TVD-like second order scheme (in a 2-D test with typical resolution). However, most of the difference is due to the use of a higher order Runge-Kutta time-stepping algorithm in the first case. If the number of sub-cycles per unit time are measured instead, then the ratio decreases to just $\approx 1.2$, and therefore high order procedures appear to be implemented in a quite efficient way. The extra coding necessary to include such routines in an existing second order code is not heavy: basically the stencils needed for reconstruction must be enlarged, and before every derivativation fluxes must be corrected with an additional high order (1-D) procedure. 

Concluding, thanks to the Eulerian approach applied to the UCT method, we have developed a unified numerical framework for MHD-like conservation laws, valid from the classical case to special/general relativistic MHD/MD, working in any set of curvilinear (even non-orthogonal) coordinates. The base scheme conserves the solenoidal constraint for the magnetic field algebraically, due to the UCT strategy, and may be extended to any formal accuracy order (for smooth solutions) with finite difference upwind reconstruction routines of different kinds. In particular, we have proposed here a limited (filtered) reconstruction based on a fifth order stencil \citep{suresh97}, which has proved to be both accurate and robust in all the tests performed.

\begin{acknowledgements}
We sincerely thanks Marco Velli, Simone Landi, Antonio Scalabrella for fruitful discussions, and an anonymous referee for the precious suggestions which have helped us to improve the scientific quality of this work. Olindo Zanotti was supported by MIUR through COFIN funds (Pacini). Niccolo' Bucciantini was supported by NASA through Hubble Fellowship Grant HST-HF-01193.01A, awarded by the Space Telescope Science Institute, which is operated by the Association of Universities for Research in Astronomy, Inc., for NASA, under contract NAS 5-26555.
\end{acknowledgements}     

\bibliographystyle{aa}
\bibliography{paper}

\begin{thebibliography}{100}
\expandafter\ifx\csname natexlab\endcsname\relax\def\natexlab#1{#1}\fi

\bibitem[{{Abramowicz} {et~al.}(1978){Abramowicz}, {Jaroszynski}, \&
  {Sikora}}]{abramowicz78}
{Abramowicz}, M., {Jaroszynski}, M., \& {Sikora}, M. 1978, \aap, 63, 221

\bibitem[{{Aloy} {et~al.}(1999){Aloy}, {Ib{\'a}{\~n}ez}, {Mart{\'{\i}}}, \&
  {M{\"u}ller}}]{aloy99}
{Aloy}, M.~A., {Ib{\'a}{\~n}ez}, J.~M., {Mart{\'{\i}}}, J.~M., \& {M{\"u}ller},
  E. 1999, \apjs, 122, 151

\bibitem[{{Anile}(1989)}]{anile89}
{Anile}, A.~M. 1989, {Relativistic fluids and magneto-fluids} (Cambdridge
  University Press)

\bibitem[{{Anninos} {et~al.}(2005){Anninos}, {Fragile}, \&
  {Salmonson}}]{anninos05}
{Anninos}, P., {Fragile}, P.~C., \& {Salmonson}, J.~D. 2005, \apj, 635, 723

\bibitem[{{Ant{\'o}n} {et~al.}(2006){Ant{\'o}n}, {Zanotti}, {Miralles},
  {Mart{\'{\i}}}, {Ib{\'a}{\~n}ez}, {Font}, \& {Pons}}]{anton06}
{Ant{\'o}n}, L., {Zanotti}, O., {Miralles}, J.~A., {et~al.} 2006, \apj, 637,
  296

\bibitem[{Arnowitt {et~al.}(1962)Arnowitt, Deser, \& Misner}]{arnowitt62}
Arnowitt, R., Deser, S., \& Misner, C.~W. 1962, The Dynamics of General
  Relativity, ed. L.~Witten (New York: John Wiley), 227--265

\bibitem[{{Balsara}(2001)}]{balsara01}
{Balsara}, D. 2001, \apjs, 132, 83

\bibitem[{{Balsara} \& {Shu}(2000)}]{balsara00}
{Balsara}, D.~S. \& {Shu}, C.-W. 2000, Journal of Computational Physics, 160,
  405

\bibitem[{{Banyuls} {et~al.}(1997){Banyuls}, {Font}, {Ib\'a\~nez},
  {Mart\'{\i}}, \& {Miralles}}]{banyuls97}
{Banyuls}, F., {Font}, J.~A., {Ib\'a\~nez}, J.~M., {Mart\'{\i}}, J.~M., \&
  {Miralles}, J.~A. 1997, \apj, 476, 221

\bibitem[{{Barmin} {et~al.}(1996){Barmin}, {Kulikovskiy}, \&
  {Pogorelov}}]{barmin96}
{Barmin}, A.~A., {Kulikovskiy}, A.~G., \& {Pogorelov}, N.~V. 1996, J. Comput.
  Phys., 126, 77

\bibitem[{{Baumgarte} \& {Shapiro}(2003)}]{baumgarte03}
{Baumgarte}, T.~W. \& {Shapiro}, S.~L. 2003, \apj, 585, 921

\bibitem[{{Blandford} \& {Znajek}(1977)}]{blandford77}
{Blandford}, R.~D. \& {Znajek}, R.~L. 1977, \mnras, 179, 433

\bibitem[{{Brio} \& {Wu}(1988)}]{brio88}
{Brio}, M. \& {Wu}, C.~C. 1988, J. Comput. Phys., 75, 400

\bibitem[{{Bucciantini} {et~al.}(2005){Bucciantini}, {Amato}, \& {Del
  Zanna}}]{bucciantini05}
{Bucciantini}, N., {Amato}, E., \& {Del Zanna}, L. 2005, \aap, 434, 189

\bibitem[{{Bucciantini} {et~al.}(2006){Bucciantini}, {Thompson}, {Arons},
  {Quataert}, \& {Del Zanna}}]{bucciantini06}
{Bucciantini}, N., {Thompson}, T.~A., {Arons}, J., {Quataert}, E., \& {Del
  Zanna}, L. 2006, \mnras, 368, 1717

\bibitem[{{Colella} \& {Woodward}(1984)}]{colella84}
{Colella}, P. \& {Woodward}, P.~R. 1984, Journal of Computational Physics, 54,
  174

\bibitem[{{De Villiers} \& {Hawley}(2003)}]{devilliers03a}
{De Villiers}, J.-P. \& {Hawley}, J.~F. 2003, \apj, 589, 458

\bibitem[{{De Villiers} {et~al.}(2003){De Villiers}, {Hawley}, \&
  {Krolik}}]{devilliers03b}
{De Villiers}, J.-P., {Hawley}, J.~F., \& {Krolik}, J.~H. 2003, \apj, 599, 1238

\bibitem[{{De Villiers} {et~al.}(2005){De Villiers}, {Hawley}, {Krolik}, \&
  {Hirose}}]{devilliers05}
{De Villiers}, J.-P., {Hawley}, J.~F., {Krolik}, J.~H., \& {Hirose}, S. 2005,
  \apj, 620, 878

\bibitem[{{Del Zanna} {et~al.}(2004){Del Zanna}, {Amato}, \&
  {Bucciantini}}]{delzanna04}
{Del Zanna}, L., {Amato}, E., \& {Bucciantini}, N. 2004, \aap, 421, 1063

\bibitem[{{Del Zanna} \& {Bucciantini}(2002)}]{delzanna02}
{Del Zanna}, L. \& {Bucciantini}, N. 2002, \aap, 390, 1177

\bibitem[{{Del Zanna} {et~al.}(2003){Del Zanna}, {Bucciantini}, \&
  {Londrillo}}]{delzanna03}
{Del Zanna}, L., {Bucciantini}, N., \& {Londrillo}, P. 2003, \aap, 400, 397

\bibitem[{{Del Zanna} {et~al.}(2006){Del Zanna}, {Volpi}, {Amato}, \&
  {Bucciantini}}]{delzanna06}
{Del Zanna}, L., {Volpi}, D., {Amato}, E., \& {Bucciantini}, N. 2006, \aap,
  453, 621

\bibitem[{{Duez} {et~al.}(2006{\natexlab{a}}){Duez}, {Liu}, {Shapiro},
  {Shibata}, \& {Stephens}}]{duez06a}
{Duez}, M.~D., {Liu}, Y.~T., {Shapiro}, S.~L., {Shibata}, M., \& {Stephens},
  B.~C. 2006{\natexlab{a}}, Physical Review Letters, 96, 031101

\bibitem[{{Duez} {et~al.}(2006{\natexlab{b}}){Duez}, {Liu}, {Shapiro},
  {Shibata}, \& {Stephens}}]{duez06b}
{Duez}, M.~D., {Liu}, Y.~T., {Shapiro}, S.~L., {Shibata}, M., \& {Stephens},
  B.~C. 2006{\natexlab{b}}, \prd, 73, 104015

\bibitem[{{Duez} {et~al.}(2005){Duez}, {Liu}, {Shapiro}, \&
  {Stephens}}]{duez05}
{Duez}, M.~D., {Liu}, Y.~T., {Shapiro}, S.~L., \& {Stephens}, B.~C. 2005, \prd,
  72, 024028

\bibitem[{{Eulderink} \& {Mellema}(1994)}]{eulderink94}
{Eulderink}, F. \& {Mellema}, G. 1994, \aap, 284, 654

\bibitem[{{Evans} \& {Hawley}(1988)}]{evans88}
{Evans}, C.~R. \& {Hawley}, J.~F. 1988, \apj, 332, 659

\bibitem[{{Font}(2003)}]{font03}
{Font}, J.~A. 2003, Living Reviews in Relativity, 6, 4

\bibitem[{{Font} \& {Daigne}(2002)}]{font02}
{Font}, J.~A. \& {Daigne}, F. 2002, \mnras, 334, 383

\bibitem[{{Font} {et~al.}(1994){Font}, {Ibanez}, {Marquina}, \&
  {Marti}}]{font94}
{Font}, J.~A., {Ibanez}, J.~M., {Marquina}, A., \& {Marti}, J.~M. 1994, \aap,
  282, 304

\bibitem[{{Gammie}(1999)}]{gammie99}
{Gammie}, C.~F. 1999, \apjl, 522, L57

\bibitem[{{Gammie} {et~al.}(2003){Gammie}, {McKinney}, \&
  {T{\'o}th}}]{gammie03}
{Gammie}, C.~F., {McKinney}, J.~C., \& {T{\'o}th}, G. 2003, \apj, 589, 444

\bibitem[{{Gammie} {et~al.}(2004){Gammie}, {Shapiro}, \& {McKinney}}]{gammie04}
{Gammie}, C.~F., {Shapiro}, S.~L., \& {McKinney}, J.~C. 2004, \apj, 602, 312

\bibitem[{{Giacomazzo} \& {Rezzolla}(2006)}]{giacomazzo06}
{Giacomazzo}, B. \& {Rezzolla}, L. 2006, J. Fluid. Mech., 562, 223

\bibitem[{{Goldreich} \& {Julian}(1969)}]{goldreich69}
{Goldreich}, P. \& {Julian}, W.~H. 1969, \apj, 157, 869

\bibitem[{{Harten} {et~al.}(1987){Harten}, {Engquist}, {Osher}, \&
  {Chakravarthy}}]{harten87}
{Harten}, A., {Engquist}, B., {Osher}, S., \& {Chakravarthy}, S. 1987, J.
  Comput. Phys., 71, 231

\bibitem[{{Harten} {et~al.}(1983){Harten}, {Lax}, \& {Van Leer}}]{harten83}
{Harten}, A., {Lax}, P.~D., \& {Van Leer}, B. 1983, SIAM Rev., 5, 1

\bibitem[{{Hawley} \& {Krolik}(2006)}]{hawley06}
{Hawley}, J.~F. \& {Krolik}, J.~H. 2006, \apj, 641, 103

\bibitem[{{Jiang} \& {Shu}(1996)}]{jiang96}
{Jiang}, G.-S. \& {Shu}, C.-W. 1996, J. Comput. Phys., 126, 202

\bibitem[{{Koide}(2003)}]{koide03}
{Koide}, S. 2003, \prd, 67, 104010

\bibitem[{{Koide} {et~al.}(2006){Koide}, {Kudoh}, \& {Shibata}}]{koide06}
{Koide}, S., {Kudoh}, T., \& {Shibata}, K. 2006, \prd, 74, 044005

\bibitem[{{Koide} {et~al.}(2000){Koide}, {Meier}, {Shibata}, \&
  {Kudoh}}]{koide00}
{Koide}, S., {Meier}, D.~L., {Shibata}, K., \& {Kudoh}, T. 2000, \apj, 536, 668

\bibitem[{{Koide} {et~al.}(1999){Koide}, {Shibata}, \& {Kudoh}}]{koide99}
{Koide}, S., {Shibata}, K., \& {Kudoh}, T. 1999, \apj, 522, 727

\bibitem[{{Komissarov}(1997)}]{komissarov97}
{Komissarov}, S.~S. 1997, Physics Letters A, 232, 435

\bibitem[{{Komissarov}(1999)}]{komissarov99}
{Komissarov}, S.~S. 1999, \mnras, 303, 343

\bibitem[{{Komissarov}(2001)}]{komissarov01}
{Komissarov}, S.~S. 2001, \mnras, 326, L41

\bibitem[{{Komissarov}(2002)}]{komissarov02}
{Komissarov}, S.~S. 2002, \mnras, 336, 759

\bibitem[{{Komissarov}(2004)}]{komissarov04a}
{Komissarov}, S.~S. 2004, \mnras, 350, 427

\bibitem[{{Komissarov}(2005)}]{komissarov05}
{Komissarov}, S.~S. 2005, \mnras, 359, 801

\bibitem[{{Komissarov}(2006{\natexlab{a}})}]{komissarov06b}
{Komissarov}, S.~S. 2006{\natexlab{a}}, \mnras, 368, 993

\bibitem[{{Komissarov}(2006{\natexlab{b}})}]{komissarov06a}
{Komissarov}, S.~S. 2006{\natexlab{b}}, \mnras, 367, 19

\bibitem[{{Komissarov} {et~al.}(2006){Komissarov}, {Barkov}, \&
  {Lyutikov}}]{komissarov06c}
{Komissarov}, S.~S., {Barkov}, M., \& {Lyutikov}, M. 2006, \mnras, 1280

\bibitem[{{Komissarov} \& {Lyubarsky}(2004)}]{komissarov04b}
{Komissarov}, S.~S. \& {Lyubarsky}, Y.~E. 2004, \mnras, 349, 779

\bibitem[{{Kozlowski} {et~al.}(1978){Kozlowski}, {Jaroszynski}, \&
  {Abramowicz}}]{kozlowski78}
{Kozlowski}, M., {Jaroszynski}, M., \& {Abramowicz}, M.~A. 1978, \aap, 63, 209

\bibitem[{{Landau} \& {Lifshitz}(1962)}]{landau62}
{Landau}, L.~D. \& {Lifshitz}, E.~M. 1962, {The classical theory of fields}
  (Oxford: Pergamon)

\bibitem[{{Leismann} {et~al.}(2005){Leismann}, {Ant{\'o}n}, {Aloy},
  {M{\"u}ller}, {Mart{\'{\i}}}, {Miralles}, \& {Ib{\'a}{\~n}ez}}]{leismann05}
{Leismann}, T., {Ant{\'o}n}, L., {Aloy}, M.~A., {et~al.} 2005, \aap, 436, 503

\bibitem[{{Lele}(1992)}]{lele92}
{Lele}, S.~K. 1992, J. Comput. Phys., 103, 16

\bibitem[{{Lichnerowicz}(1967)}]{lichnerowicz67}
{Lichnerowicz}, A. 1967, Relativistic hydrodynamics and magnetohydrodynamics
  (New York: Benjamin)

\bibitem[{{Liu} \& {Osher}(1998)}]{liu98}
{Liu}, X.-D. \& {Osher}, S. 1998, J. Comput. Phys., 141, 1

\bibitem[{{Londrillo} \& {Del Zanna}(2000)}]{londrillo00}
{Londrillo}, P. \& {Del Zanna}, L. 2000, \apj, 530, 508

\bibitem[{{Londrillo} \& {Del Zanna}(2004)}]{londrillo04}
{Londrillo}, P. \& {Del Zanna}, L. 2004, J. Comput. Phys., 195, 17

\bibitem[{{Mart{\'{\i}}} \& {M{\"u}ller}(1996)}]{marti96}
{Mart{\'{\i}}}, J.~M. \& {M{\"u}ller}, E. 1996, J. Comput. Phys., 123, 1

\bibitem[{{Mart{\'{\i}}} \& {M{\"u}ller}(2003)}]{marti03}
{Mart{\'{\i}}}, J.~M. \& {M{\"u}ller}, E. 2003, Living Reviews in Relativity,
  6, 7

\bibitem[{{McKinney}(2005)}]{mckinney05}
{McKinney}, J.~C. 2005, \apjl, 630, L5

\bibitem[{{McKinney}(2006{\natexlab{a}})}]{mckinney06a}
{McKinney}, J.~C. 2006{\natexlab{a}}, \mnras, 367, 1797

\bibitem[{{McKinney}(2006{\natexlab{b}})}]{mckinney06b}
{McKinney}, J.~C. 2006{\natexlab{b}}, \mnras, 368, 1561

\bibitem[{{McKinney}(2006{\natexlab{c}})}]{mckinney06c}
{McKinney}, J.~C. 2006{\natexlab{c}}, \mnras, 368, L30

\bibitem[{{McKinney} \& {Gammie}(2004)}]{mckinney04}
{McKinney}, J.~C. \& {Gammie}, C.~F. 2004, \apj, 611, 977

\bibitem[{{Michel}(1972)}]{michel72}
{Michel}, F.~C. 1972, \apss, 15, 153

\bibitem[{{Mignone} \& {Bodo}(2006)}]{mignone06}
{Mignone}, A. \& {Bodo}, G. 2006, \mnras, 368, 1040

\bibitem[{{Mignone} {et~al.}(2005){Mignone}, {Plewa}, \& {Bodo}}]{mignone05}
{Mignone}, A., {Plewa}, T., \& {Bodo}, G. 2005, \apjs, 160, 199

\bibitem[{{Misner} {et~al.}(1973){Misner}, {Thorne}, \& {Wheeler}}]{misner73}
{Misner}, C.~W., {Thorne}, K.~S., \& {Wheeler}, J.~A. 1973, {Gravitation} (San
  Francisco: Freeman)

\bibitem[{{Mizuno} {et~al.}(2004){Mizuno}, {Yamada}, {Koide}, \&
  {Shibata}}]{mizuno04}
{Mizuno}, Y., {Yamada}, S., {Koide}, S., \& {Shibata}, K. 2004, \apj, 615, 389

\bibitem[{{Myong} \& {Roe}(1998)}]{myong98}
{Myong}, R.~S. \& {Roe}, P.~L. 1998, J. Comput. Phys., 147, 545

\bibitem[{{Nishikawa} {et~al.}(2005){Nishikawa}, {Richardson}, {Koide},
  {Shibata}, {Kudoh}, {Hardee}, \& {Fishman}}]{nishikawa05}
{Nishikawa}, K.-I., {Richardson}, G., {Koide}, S., {et~al.} 2005, \apj, 625, 60

\bibitem[{{Noble} {et~al.}(2006){Noble}, {Gammie}, {McKinney}, \& {Del
  Zanna}}]{noble06}
{Noble}, S.~C., {Gammie}, C.~F., {McKinney}, J.~C., \& {Del Zanna}, L. 2006,
  \apj, 641, 626

\bibitem[{{Papadopoulos} \& {Font}(1998)}]{papadopoulos98}
{Papadopoulos}, P. \& {Font}, J.~A. 1998, \prd, 58, 024005

\bibitem[{{Pons} {et~al.}(1998){Pons}, {Font}, {Ibanez}, {Marti}, \&
  {Miralles}}]{pons98}
{Pons}, J.~A., {Font}, J.~A., {Ibanez}, J.~M., {Marti}, J.~M., \& {Miralles},
  J.~A. 1998, \aap, 339, 638

\bibitem[{{Roe}(1981)}]{roe81}
{Roe}, P.~L. 1981, J. Comput. Phys., 43, 357

\bibitem[{{Ryu} {et~al.}(2006){Ryu}, {Chattopadhyay}, \& {Choi}}]{ryu06}
{Ryu}, D., {Chattopadhyay}, I., \& {Choi}, E. 2006, \apjs, 166, 410

\bibitem[{{Shibata} {et~al.}(2006){Shibata}, {Duez}, {Liu}, {Shapiro}, \&
  {Stephens}}]{shibata06}
{Shibata}, M., {Duez}, M.~D., {Liu}, Y.~T., {Shapiro}, S.~L., \& {Stephens},
  B.~C. 2006, Physical Review Letters, 96, 031102

\bibitem[{{Shibata} \& {Sekiguchi}(2005)}]{shibata05}
{Shibata}, M. \& {Sekiguchi}, Y.-I. 2005, \prd, 72, 044014

\bibitem[{{Shu}(1997)}]{shu97}
{Shu}, C.-W. 1997, NASA ICASE Rep., 97, 65

\bibitem[{{Shu} \& {Osher}(1988)}]{shu88}
{Shu}, C.-W. \& {Osher}, S. 1988, J. Comput. Phys., 77, 439

\bibitem[{Sloan \& Smarr(1985)}]{sloan85}
Sloan, J. \& Smarr, L.~L. 1985, General relativistic magnetohydrodynamics, ed.
  J.~L. J.~Centrella \& R.~Bowers (Boston: Jones and Bartlett), 52--68

\bibitem[{{Smarr} \& {York}(1978)}]{smarr78}
{Smarr}, L. \& {York}, Jr., J.~W. 1978, \prd, 17, 2529

\bibitem[{{Spitkovsky}(2006)}]{spitkovsky06}
{Spitkovsky}, A. 2006, \apjl, 648, L51

\bibitem[{{Suresh} \& {Huynh}(1997)}]{suresh97}
{Suresh}, A. \& {Huynh}, H.~T. 1997, J. Comput. Phys., 136, 83

\bibitem[{{Takahashi} {et~al.}(1990){Takahashi}, {Nitta}, {Tatematsu}, \&
  {Tomimatsu}}]{takahashi90}
{Takahashi}, M., {Nitta}, S., {Tatematsu}, Y., \& {Tomimatsu}, A. 1990, \apj,
  363, 206

\bibitem[{{Thorne} \& {MacDonald}(1982)}]{thorne82}
{Thorne}, K.~S. \& {MacDonald}, D. 1982, \mnras, 198, 339

\bibitem[{{Torrilhon}(2004)}]{torrilhon04}
{Torrilhon}, M. 2004, Journal of Plasma Physics, 69, 253

\bibitem[{{van~Putten}(1993)}]{vanputten93}
{van~Putten}, M.~H.~P.~M. 1993, J. Comput. Phys., 99, 341

\bibitem[{{Wald}(1974)}]{wald74}
{Wald}, R.~M. 1974, \prd, 10, 1680

\bibitem[{{Weber} \& {Davis}(1967)}]{weber67}
{Weber}, E.~J. \& {Davis}, L.~J. 1967, \apj, 148, 217

\bibitem[{Weinberg(1972)}]{weinberg72}
Weinberg, S. 1972, Gravitation and Cosmology (New York: Wiley)

\bibitem[{{Wilson} \& {Mathews}(2003)}]{wilson03}
{Wilson}, J.~R. \& {Mathews}, G.~J. 2003, {Relativistic Numerical
  Hydrodynamics} (Cambridge, UK: Cambridge University Press)

\bibitem[{{Yee}(1966)}]{yee66}
{Yee}, K. 1966, IEEE Trans. Antennas Propag., 14, 302

\bibitem[{{York}(1979)}]{york79}
{York}, Jr., J.~W. 1979, Kinematics and dynamics of general relativity, ed.
  L.~Smarr (Cambridge: Cambridge Univ. Press), 83--126

\bibitem[{{Zhang}(1989)}]{zhang89}
{Zhang}, X.-H. 1989, \prd, 39, 2933

\end{thebibliography}


\appendix
\section{Finite difference procedures}

ECHO employs finite difference piecewise polynomial high order procedures for interpolation, reconstruction, and derivation. Compact-like (implicit) routines are also implemented, but we do not discuss them here. Below we will indicate with $n$ the order of accuracy of the single procedures, while $r$ will retain the meaning of the spatial accuracy of the overall scheme. 

\subsection{Interpolation (INT)}
\label{sect:interp}

Interpolation is explicitly needed to approximate the magnetic field components at step 1 in Sect.~\ref{sect:discr}, but it also provide the building blocks for upwind reconstruction methods. For any kind of polynomial interpolation, it is convenient to calculate the coefficients by means of the Lagrange formula. For a stencil of $n$ points $x_i$ (either cell centers or intercell points), the polynomial approximating a function $f(x)$ to $n$-th order is
\be
\label{eq:lagrange}
p_n(x)=\sum_{i=1}^{n}a_if_i,~~~a_i=\prod_{k=1,\,k\neq i}^{n}\frac{x-x_k}{x_i-x_k},
\ee
where by construction $p_n(x_i)=f_i\equiv f(x_i)$. For the case of magnetic field interpolation, we need to approximate a function $f(x)$ at cell center $x_j$ for given intercell values $f_{j+1/2}$. Application of Eq.~(\ref{eq:lagrange}) to a symmetric stencil around $x_j$ gives the expressions
\bea
f_j & = & (f_{j-1/2}+f_{j+1/2})/2, \\
f_j & = & (-f_{j-3/2}+9f_{j-1/2}+9f_{j+1/2}-f_{j+3/2})/16, \\
f_j & = & (3f_{j-5/2}-25f_{j-3/2}+150f_{j-1/2}+ \nonumber \\
    &   & +150f_{j+1/2}-25f_{j+3/2}+3f_{j+5/2})/256,
\eea
respectively for $n=2$, $n=4$, and $n=6$. Thus, the $n$-th order formula should be used for an overall scheme with $r\leq n$.

\subsection{Reconstruction (REC)}
\label{sect:rec}

The reconstruction process employed in ECHO is again an operation based on piecewise polynomial interpolation. Given a stencil of $n$ grid points $\{x_j\}$ (cell centers) with corresponding values $\{f_j\}$ of the discretized function $f(x)$ (in ECHO the primitive variables, see step 2), the problem is to find a $n$-th order approximation of the intercell value $f_{j+1/2}$. Note that in the numerical literature high order reconstruction is usually implemented to find directly the $\hat{f}_{j+1/2}$ numerical fluxes of step 4 (called reconstruction via the primitive function), corresponding to our REC+DER combined operations. Therefore, the polynomial coefficients presented here will differ with those usually found in the literature. Contrary to the centered interpolations seen above, in shock-capturing schemes \emph{upwind} interpolation is needed, that is based on either left-biased ($L$) or right-biased ($R$) stencils. To achieve this, typically $n$ is chosen an odd number and the two stencils are taken symmetric with respect to $x_{j+1/2}$. Moreover, the same reconstruction routine $\mathcal{R}(\{f_j\})$ may be employed for both $L$ and $R$ procedures:
\bea
f_{j+1/2}^L & = & \mathcal{R}(f_{j-(n-1)/2},\ldots,f_{j+(n-1)/2}),\\
f_{j+1/2}^R & = & \mathcal{R}(f_{j+1+(n-1)/2},\ldots,f_{j+1-(n-1)/2}).
\eea
For $n=1$ we have the expected upwind constant approximations $f_{j+1/2}^L=f_j$, $f_{j+1/2}^R=f_{j+1}$. For $n>1$ we have to face the problem that the two stencils may contain a discontinuity, hence sub-stencils should be used in order to avoid Gibbs oscillations and the above formula actually refers to the \emph{optimal} stencils providing an order $n$ only for smooth solutions. 

For $n=3$ we have quadratic interpolation. By applying Eq.~(\ref{eq:lagrange}), the left fixed-stencil reconstruction (only left reconstructions will be considered hereafter) based on the optimal stencil is
\be
\label{eq:opt3}
f_{j+1/2}=(-f_{j-1}+6f_{j}+3f_{j+1})/8.
\ee
In TVD-like reconstructions, based on the same $n=3$ stencil used above, third order is sacrificed for sake of stability by resorting to second order for continuos fields and to first order when a discontinuity is present. These schemes are based on piecewise linear reconstruction and monotonicity is typically enforced by making use of slope limiters
\be
f_{j+1/2}=f_j+\textstyle{\frac{1}{2}}S(\Delta_-f_j,\Delta_+f_j),
\ee
where $\Delta_\pm f_j=\pm (f_{j\pm 1} - f_j)$ and the slope $S$ can be for example the \emph{MinMod} (MM2 in ECHO) limiter
\be
\mathrm{mm}(x,y)=\textstyle{\frac{1}{2}}[\mathrm{sgn}(x)+\mathrm{sgn}(y)]\mathrm{min}(|x|,|y|),
\ee
or the so-called \emph{Monotonized Centered} (MC2 in ECHO) limiter
\be
\mathrm{mc}(x,y)=\textstyle{\frac{1}{2}}[\mathrm{sgn}(x)+\mathrm{sgn}(y)]\mathrm{min}(2|x|,2|y|,\textstyle{\frac{1}{2}}|x+y|).
\ee
Usually reconstruction based on MM2 is safer but more smearing, while MC2 provides a good compromise between robustness and accuracy. Note that at local (smooth) extrema all limited reconstructions of this kind drop to first order. ENO schemes follow a different strategy. In ENO2, one between the two linear interpolations based on 2-point sub-stencils
\bea
f_{j+1/2}^{\,(1)} & = f_j+\textstyle{\frac{1}{2}}\Delta_-f_j  & = (-f_{j-1}+3f_{j})/2, \\
f_{j+1/2}^{\,(2)} & = f_j+\textstyle{\frac{1}{2}}\Delta_+f_j  & = (f_{j}+f_{j+1})/2,
\eea
is chosen, with selection procedures based on smoothness criteria to ensure the (essentially) non-oscillatory behavior. Thus, ENO2 always employs a piecewise linear interpolation. The possibility to achieve the optimal third order reconstruction of Eq.~(\ref{eq:opt3}) is provided by the weighting process in the WENO3 procedure:
\be
f_{j+1/2}=\omega_1 f_{j+1/2}^{\,(1)} + \omega_2 f_{j+1/2}^{\,(2)},
\ee
where the optimal reconstruction is found for $\omega_1=1/4$ and $\omega_2=3/4$. In the (nonlinear) selection process these are the limits for smooth fields, otherwise a different combination (resulting in a lower order) is achieved and for discontinuous fields WENO3 is equivalent to ENO2.

Analogous possibilities for ENO-like schemes are offered by reconstruction based on the $n=5$ stencil. The optimal choice yielding fifth order accuracy is
\be
\label{eq:opt5}
f_{j+1/2}=(3f_{j-2} -20f_{j-1} +90f_j +60f_{j+1} -5f_{j+2})/128,
\ee
while the three 3-point sub-stencils provide the quadratic interpolations
\bea
f_{j+1/2}^{\,(1)} & = & (3f_{j-2}-10f_{j-1}+15f_{j})/8, \\
f_{j+1/2}^{\,(2)} & = & (-f_{j-1}+6f_{j}+3f_{j+1})/8, \\
f_{j+1/2}^{\,(3)} & = & (3f_{j}+6f_{j+1}-f_{j+2})/8,
\eea
which are easily obtained as usual by either use of Eq.~(\ref{eq:lagrange}) or by Taylor expansion. In ENO3 third order reconstruction is always obtained by choosing the smoothest among the above interpolations. In WENO5 the combination
\be
f_{j+1/2}=\omega_1 f_{j+1/2}^{\,(1)} + \omega_2 f_{j+1/2}^{\,(2)} + \omega_3 f_{j+1/2}^{\,(3)},
\ee
is used, and the optimal fifth order reconstruction in Eq.~(\ref{eq:opt5}) is retrieved when $\omega_1=1/16$, $\omega_2=10/16$ and $\omega_3=5/16$, obtained for smooth fields. Another possibility is provided by the CENO3 algorithm \citep{liu98}, which is basically equivalent to ENO3 for smooth fields (thus both achieve third order at most) and it reduces to lower order TVD reconstruction (hence even to first order) in the presence of discontinuities, but not at smooth extrema. The robustness and accuracy of this scheme were comprehesively tested in Paper~I and II. 

A different strategy is followed by MP (\emph{Monotonicity Preserving}) methods \citep{suresh97}: first the high order reconstruction, like that in Eq.~(\ref{eq:opt5}) for MP5, is constructed, then, if spurious oscillations are found, a nonlinear filter based on limiting algorithms is applied to reduce them, retrieving first order approximations only where needed (like in CENO). An approach similar to MP is that followed in the celebrated PPM \citep[\emph{Piecewise Parabolic Method, }][]{colella84}, very popular among astrophysicists, due to the rather sharp profiles provided at discontinuities, and used in special relativistic HD and MHD too \citep{marti96,mignone05,leismann05}. However, that method has the drawback of reducing to first order even at smooth extrema, just like TVD. Moreover, the post-processing filters for PPM are rather involved and heavily system-dependent(especially the steepening of contact-like discontinuities), thus in conflict with the philosophy adopted here. On the other hand, MP methods are particularly suitable for component-wise reconstruction and these filters can be applied to a variety of explicit interpolants, to higher order WENO methods \citep{balsara00}, or even to compact interpolations with spectral-like resolution \citep{lele92}. The MP5 algorithm based on the $n=5$ explicit reconstruction of Eq.~(\ref{eq:opt5}) has been shown here to be both highly accurate and robust in all tests, and we thus recommend its use. We refer to the original paper for a description of the nonlinear filter.

\subsection{Derivation (DER)}
\label{sect:der}

The derivation operation was encountered at step 4 to provide the numerical flux function $\hat{f}_{j+1/2}$, given a stencil of intercell fluxes $\{f_{j+1/2}\}$. This must be done in such a way that $(\hat{f}_{j+1/2}-\hat{f}_{j-1/2})/h$ is an appropriate high order approximation of the $f^\prime (x)$ first derivative calculated at $x=x_j$, where $h$ is the (constant) grid spacing. Let us then start by looking for a finite difference approximation of the first derivative. It is convenient to write it as
\bea
hf^\prime (x_j) & \approx & \hat{f}_{j+1/2} - \hat{f}_{j-1/2} = a(f_{j+1/2}-f_{j-1/2}) +
\nonumber \\
 & & + b(f_{j+3/2}-f_{j-3/2})+c(f_{j+5/2}-f_{j-5/2}),
\label{eq:hf1}
\eea
where we have truncated the approximation up to sixth order. If we now expand both sides of the above equation in Taylor series around $x_j$ we find the system
\be
hf^{\,(1)}_j \!=\!\sum_{k=0}^{\infty}f_j^{\,(k)}\frac{h^k}{k!2^k}[1-(-1)^k][a+3^k b+5^k c],
\ee
where the exponents indicate derivation of the corresponding order and where clearly all terms with even $k$ vanish. For $n=2$, where $b=c=0$, we simply find $a=1$. For $n=4$, where only $c=0$, the above system is readily solved by $a=9/8$, $b=-1/24$. Finally, for $n=6$ the solution is $a=75/64$, $b=-25/384$, $c=3/640$.
The next step is to write
\be
\label{eq:fder}
\hat{f}_{j+1/2}= \! d_0f_{j+1/2}+d_2(f_{j-1/2}+f_{j+3/2})+d_4(f_{j-3/2}+f_{j+5/2}),
\ee
and comparison with Eq.~(\ref{eq:hf1}) provides the relations $d_0=a+b+c$, $d_2=b+c$, $d_4=c$. For $n=2$ $d_0=1$, $d_2=d_4=0$ and $\hat{f}_{j+1/2}=f_{j+1/2}$, as expected. Thus, no extra high order corrections on numerical fluxes are needed for schemes up to second order. For $n=4$ we find $d_0=13/12$, $d_2=-1/24$, and $d_4=0$. Finally, for $n=6$ we find $d_0=1067/960$, $d_2=-29/480$, and $d_4=3/640$. 

In order to highlight the nature of the DER procedure as a correction for higher than second order approximations, it is convenient to rewrite Eq.~(\ref{eq:fder}) in the form
\be
\hat{f}_{j+1/2}=f_{j+1/2}-\frac{1}{24}\Delta^{(2)} f_{j+1/2}+\frac{3}{640}\Delta^{(4)} f_{j+1/2},
\ee
where only the first term is retained for $n=2$, the second is introduced for $n=4$, and the complete expression is used for $n=6$. For a generic index $i$ the second and fourth order numerical derivatives are respectively given by
\bea
\Delta^{(2)} f_{i} & = & f_{i-1}-2f_{i}+f_{i+1}, \\
\Delta^{(4)} f_{i} & = & \Delta^{(2)} f_{i-1}-2\Delta^{(2)} f_{i}+\Delta^{(2)} f_{i+1} = \nonumber \\
& = & f_{i-2}-4f_{i-1}+6f_{i}-4f_{i+1}+f_{i+2}.
\eea
Notice that here only DER operators based on centered, symmetric stencils have been considered. The high order corrections described above can be easily turned into non-oscillatory algorithms by any sort of limiting or stencil selection upwind process, like those employed for REC.

\end{document}